\documentclass[5p]{elsarticle}

\usepackage[hidelinks]{hyperref}

\journal{Ultramicroscopy}

\usepackage{amstext}

\usepackage{graphicx}

\begin{document}

\begin{frontmatter}

\title{Reconstructing the exit wave in high-resolution transmission
  electron microscopy using machine learning}

\author[camd]{Matthew Helmi Leth Larsen}
\author[camd]{Frederik Dahl}
\author[topsoe]{Lars P. Hansen}
\author[berk]{Bastian Barton}
\author[berk]{Christian Kisielowski}
\author[vision]{Stig Helveg}
\author[comp]{Ole Winther}
\author[nlab]{Thomas W. Hansen}
\author[camd]{Jakob Schiøtz\corref{corr}}

\address[camd]{Computational Atomic-scale Materials Design (CAMD), Department of Physics,
  Technical University of Denmark, DK-2800 Kgs.\  Lyngby, Denmark}
\address[topsoe]{Haldor Topsøe A/S, Haldor Topsøes Allé 1, DK-2800 Kgs.\ 
  Lyngby, Denmark}
\address[berk]{The Molecular Foundry, Lawrence Berkeley National Laboratory,
  One Cyclotron Road, CA 94720 Berkeley, USA}
\address[vision]{Center for Visualizing Catalytic Processes (VISION), Department of Physics,
  Technical University of Denmark, DK-2800 Kgs.\  Lyngby, Denmark}
\address[comp]{Department of Applied Mathematics and Computer Science,
  Technical University of Denmark, DK-2800 Kgs.\  Lyngby, Denmark}
\address[nlab]{National Center for Nano Fabrication and Characterization,
  Technical University of Denmark, DK-2800 Kgs.\  Lyngby, Denmark}
\cortext[corr]{Corresponding author: schiotz@fysik.dtu.dk}

\begin{abstract}
  Reconstruction of the exit wave function is an important route to
  interpreting high-resolution transmission electron microscopy
  (HRTEM) images.  Here we demonstrate that convolutional neural
  networks can be used to reconstruct the exit wave from a short focal
  series of HRTEM images, with a fidelity comparable to
  conventional exit wave reconstruction.  We use a fully convolutional
  neural network based on the U-Net architecture, and demonstrate that
  we can train it on simulated exit waves and simulated HRTEM images
  of graphene-supported molybdenum disulphide (an industrial
  desulfurization catalyst).  We then apply the trained network to
  analyse experimentally obtained images from similar samples, and
  obtain exit waves that clearly show the atomically resolved
  structure of both the MoS$_2$ nanoparticles and the graphene
  support.  We also show that it is possible to successfully train the
  neural networks to reconstruct exit waves for 3400
  different two-dimensional materials taken from the Computational 2D
  Materials Database of known and proposed two-dimensional materials.
\end{abstract}


\begin{keyword}
{exit wave reconstruction},
{HRTEM},
{2D Materials},
{Machine Learning}
\end{keyword}

\end{frontmatter}


\section{Introduction}
\label{sec:introduction}

Machine learning has become a powerful tool for analyzing images. In
fact, machine learning is a nascent tool in electron microscopy that
is envisioned to have a large potential for quantitative image analysis
\cite{Kalinin:2019ds,Spurgeon:2020jr}. In electron microscopy,
applications of machine learning have up to now included
segmentation of medical images \cite{Quan2016}, grain and phase
identification \cite{Holm:2020fz,Azimi:2018cy,DeCost:2019ix}, noise
filtering \cite{Lin2021:TEMImageNet,Vincent2021:Denoise} and in-plane location of atoms
\cite{Madsen:2018ey,Ziatdinov:2017ct,Lee:2020bk}. Moreover, Ede
\emph{et al.} showed recently that the imaginary part of the exit wave
function can be reconstructed from the real part using a
convolutional neural network \cite{2020arXiv200110938E} and Meyer
showed that off-axis holograms, where phase information is recorded
directly into the image, can be reconstructed using neural networks
\cite{Meyer:2008je}. In this work we suggest that neural networks
could potentially solve the classical phase problem and thus retrieve
the entire electron wave function exiting the specimen in a transmission
electron microscopy experiment.

Aberration-corrected high-resolution transmission electron microscopy
(HRTEM) is one of the important experimental techniques to study the
structure of materials at the atomic scale.  The maximal amount of
information about the sample is present in the exit wave, i.e.\ the
wavefunction of the electrons exiting the sample.  As the image
is formed, some of this information is lost, both due to aberration in
the lenses, and because the camera 
detects the intensity of the wave, not its phase.

It is well established that the full exit wave can be reconstructed
from a focal series of images
\cite{deBeeck:1996ii,Tiemeijer:2012em}.  A series of typically around
20--50 images with varying defocus is used to numerically reconstruct
the most likely wave function of the electron beam as it exits the
sample.  This can then be used to further reconstruct information about
the chemical composition and 3D structure of the sample
\cite{Chen:2015ik,Chen2021exitwave}.  For beam-sensitive samples
\cite{VanDyck:2015fg}, exit wave reconstruction has the
advantage of being averaging in nature such that information from many images with very low
signal-to-noise ratio is combined in a single exit wave image of
superior signat-to-noise ratio
\cite{Chen2021exitwave}.  Several numerical algorithms are available
for reconstructing the exit wave \cite{Thust1996exitwave,Hsieh:2004ca,Allen:2004hc}.

Here we examine a Convolutional Neural Network (CNN) as an
alternative way to reconstruct the exit wave.  This
reconstruction is possible from a low number of HRTEM images, and
with the advantage that the detailed knowledge of the aberration parameters of the
microscope is not needed.
We envision that this can be developed into a tool for on-the-fly exit
wave reconstruction while taking data on the microscope, perhaps
supplemented with more traditional exit wave reconstruction as post
processing.  In the present case, the images were convoluted with the
effects of defocus, first order astigmatism, coma, and blurring including
focal spread.  In this case a focal series of two to three simulated
HRTEM images were
sufficient to reconstruct the
exit wave with sufficient accuracy in order to extract quantitative
information about the sample.  In principle, it should be
straightforward to extend the present method to situations with low
signal-to-noise ratio and more
unknown aberrations, in which case it is likely that a larger focal
series will be needed.

Recently, atomically thin two-dimensional (2D) materials have been an
active topic of research, with applications ranging from electronics
to energy storage and catalysis
\cite{Ferrari:2015co,Bhimanapati:2015bo}.  For example, molybdenum
disulphide (MoS$_2$) is the preferred catalyst for removing sulphur
from crude oil, and is one of the reasons that acid rain is no longer
one of the most pressing environmental problems \cite{Chorkendorff:2007book}.
In this paper, we focus on exit wave reconstruction for the rapidly
growing class of 2D materials, although the methods should be
generally applicable.  We show that neural networks can reconstruct
the exit wave both when trained to a single material, and to a
database of thousands of proposed 2D materials.  The reconstruction is
of sufficient quality to permit analysis of the image peaks associated
with the atomic columns e.g.\ by using
Argand plots to identify the type and number of elements in the material
\cite{Chen:2015ik}.

We also show that it is possible to train the
neural network purely on simulated data, and apply it successfully to
experimental images of non-trivial complexity, in this case a model
catalyst based on molybdenum disulphide.

\section{Methods}
\label{sec:methods}

The neural network architecture is a Unet \cite{Ronneberger:2015gk} /
FusionNet \cite{Quan2016} architecture, very close to the one used by
Madsen \emph{et al.} \cite{Madsen:2018ey}, with the main modification
that concatenation is used instead of elementwise addition for the
skip connections.  A linear activation function is applied in the
output layer, as exit wave reconstruction is a regression problem
rather than a classification/segmentation problem.  Details of the
architecture can be found in the Supplementary Online Information
(SOI Sec.~S1).  The neural network is implemented
and trained using the Keras interface \cite{Chollet:2018uk} to
Tensorflow version 2.5 \cite{tensorflow}. We train using simulated
images only.  We computer-generate a training set and a corresponding
validation set of atomic structures, using the Atomic Simulation
Environment (ASE) \cite{HjorthLarsen:2017hn}.

Three data sets of increasing complexity were created.  The first
consists of nanoparticles (nanoflakes) of molybdenum disulphide
(MoS$_2$).  In this data set we ignore that the nanoparticles will
typically be supported on another material in the microscope.
Nevertheless this data set will be relevant for e.g.\ edges of MoS$_2$
films on a TEM grid, where no support is visible in the region of interest.

The second dataset is 
MoS$_2$ supported on a graphene substrate.  A
nanoflake of graphene and one of MoS$_2$ are generated in the
computer, and are placed with a random distance between 3.3 and 7.0
\AA.  One quarter of the cases are placed with the lattice vectors of
the two layers in the same directions, another quarter with a rotation
of 15$^\circ$, one quarter with a rotation of 30$^\circ$, and the rest
with a random rotation.  In both of these datasets 1000 samples are
created for training, and 1000 for validation.

The third dataset consists of nanoflakes of materials from the
Computational 2D-materials Database (C2DB) \cite{Haastrup:2018ca} in
the latest version dated 2021/06/24.  This version of the database
contains 4056 known or proposed 2D materials, but a significant number
of these have very complex structures where the quasi-2D material
contains a large number of atomic layers.  We filtered the database so
we only keep structures at most eight atoms in the unit cell, that
left us with 3393 materials.  Two samples are created of each
material.  Materials are randomly assigned to the
training or validation set with a probability of 2:1, but in such a
way that all materials containing the same set of elements are
assigned to the same set.

For all three datasets, vacancies and holes are introduced
in the systems.  A vacancy is introduced by selecting a random atom
and removing it; holes are made by selecting a random atom and then
removing the entire atomic column.  In the case of MoS$_2$, if a sulphur
atom is selected then a vacancy would be removing just that atom,
whereas creating a hole would be removing an S$_2$ dimer.  If a
molybdenum atom is selected there will be no difference.  We select
5\% of the atoms for vacancy creation, then 5\% for hole creation.
All atomic positions are then perturbed by adding a Gaussian with
mean of 0 and spread of 0.01\AA{} to all atomic positions.  Finally, all
samples are tilted by a random angle up to 10$^\circ$ in a random
direction.

Exit waves are then calculated using the multislice algorithm
\cite{Goodman:1974ku,kirkland2010book}, using the abTEM software
\cite{Madsen2021:abtem}.  The lateral sampling of the wave function is
0.05 Å, and the slice thickness is 0.2 Å, see the SOI Sec.~S2.  As a simple model of
atomic vibrations, the potential of the atoms is smeared by a Gaussian
with with
\begin{equation}
  \label{eq:2}
  \left<u^2\right> = {3 \hbar^2 \over 2 m k_B \theta_D} \cdot \left(\frac14 + {T
      \over \theta_D}\right)
\end{equation}
where $m$ is the atomic mass and $\theta$ is the Debye temperature
\cite[supplementary online information]{Mannebach:2015}.
As the same value must be used for all atoms, we use the atomic mass
of Sulphur.  With $\theta_D = 580$ K for bulk MoS$_2$
\cite{Mannebach:2015}, this gives a value of $\left<u^2\right> =
0.0030 \mbox{\AA}^2$ at 300 K.  Our own \emph{ab initio} molecular dynamics
simulations of MoS$_2$ gives a somewhat larger value, which is
expected as molecular dynamics ignores the quantization of the phonons
which is important below the Debye temperature.  As an approximation,
we also use this value of $\left<u^2\right>$ for the materials in the
C2DB. If the reconstructed exit wave is to be used to gain
information about the vibrational amplitudes of different kinds of
atoms, as is done in Ref.~\cite{Chen2021exitwave}, the phonons
need to be modelled with a more sophisticated method, such as the
frozen phonon method, at a significant cost in computational burden
(up to two orders of magnitude).

After generating the exit waves, the abTEM software is used to generate
typically three images of the sample by applying a Contrast Transfer
Function (CTF), Poisson noise in the detector, and a Modulation
Transfer Function (MTF) introducing correlations in the noise.  This
is described in detail elsewhere \cite{Madsen:2018ey}.  The
parameters of the CTF and the MTF (collectively referred to as the
``microscope parameters'') are drawn from distributions given in Table
\ref{tab:parameters}.  The three images have the same microscopy
parameters except that the defocus is changed by $5 \pm 0.1$ nm
between the images.  If a different number of images is used, the
total variation in defocus remains at 10 nm.
\begin{table}
  \centering
  \begin{tabular}{lcc}
    \hline\hline
    Parameter & Lower bound & Upper bound \\
    \hline
    Acceleration voltage & \multicolumn{2}{c}{50 keV} \\
    Defocus ($\Delta f$) & -150 \AA & 150 \AA\\
    Spherical aberration  ($C_s$) & -15 $\mu$m & 15 $\mu$m \\
    2-fold order astigmatism& ~ & ~ \\
    ~~~amplitude  ($C_{12}$) & 0 & 25 \AA \\
    ~~~angle & 0 & $2\pi$ \\
    Coma & ~ & ~ \\
    ~~~amplitude  ($C_{21}$) & 0 & 600 \AA \\
    ~~~angle & 0 & $2\pi$ \\
    Focal spread & 5 \AA & 20 \AA \\
    Blur & 0.5 \AA & 1.5 \AA \\
    Electron dose & $10^{2.5} \text{\AA}^{-2}$ 
                            & $10^{5.0} \text{\AA}^{-2}$ \\
    Resolution & 0.10 \AA & 0.11 \AA\\
    MTF $c_1$ & -0.6 & 0.2 \\
    MTF $c_2$ & 0.1 & 0.2 \\
    MTF $c_3$ & 0.6 & 1.8 \\
    \hline\hline
  \end{tabular}
  \caption{Microscope parameters.  For each image series, a set of
    microscope parameters are drawn within the limits given here,
    except the acceleration voltage which is kept constant.
    All distributions are uniform, except for the dose which is
    exponential.  The defocus of the first image is picked so the
    defocus of all images are within the bound specified.}
  \label{tab:parameters}
\end{table}

The expensive part of the image simulation is the multislice algorithm
calculating the interaction between the electron beam and the sample.
The action of the CTF and the MTF are computationally cheap, and for
that reason it is convenient to generate multiple images of the same
sample with varying microscope parameters.  Depending on the
computational setup, it may be most convenient to generate images
on-the-fly during training, such that the network sees different
images of the same samples in each training epoch, or it is possible
to pre-generate and store the images.  In this work we pre-generated ten
epochs of images for the training set, and one for the validation
set.  We then cycled through the pre-generated epochs for the actual
training, which were up to 200 epochs (leading to each image being
reused 20 times).

The neural network is trained using the mean square error (MSE) as the
loss function, with the RMSprop training algorithm as implemented in
Keras, and a learning rate of $5 \times 10^{-4}$.  We also tried using
the Adam algorithm \cite{kingma2017adam}, and saw similar but slightly
less stable results, whereas Adam with the AMSgrad modification gave
almost identical results to RMSprop.  Increasing the learning rate
above $1 \times 10^{-3}$ would make the training unstable, and decreasing
it below $5 \times 10^{-4}$ was detrimental to the learning.  Training
curves showing the loss function of the training and validation set
are shown in the SOI (Fig.~S2).  In spite of the
reuse of pre-generated images, the training curves do not show signs
of overfitting.  We therefore did not use regularization in the neural
network.

The sharp potential of the nucleus causes some amount of annular
structures to appear in the exit wave, in spite of the application of
Debye-Waller smearing.  This fine structure contain little or no
information of value when analysing the exit waves.  However, the neural network
will attempt to recover this structure, leading to an overall small
degradation of its ability to recover more important information about
the main peaks associated with the atomic columns.  For simplicity, we have filtered the exit waves prior to
training by folding them with a Gaussian with a spread of 15 pm, see
SOI Fig.~S4.  This leads to a significant improvement
in the network performance, in particular when it comes to extracting
quantitative information from the peak values.

\section{Results and discussion}
\label{sec:results-discussion}

\begin{figure*}[t]
  \centering
  \includegraphics[width=\linewidth]{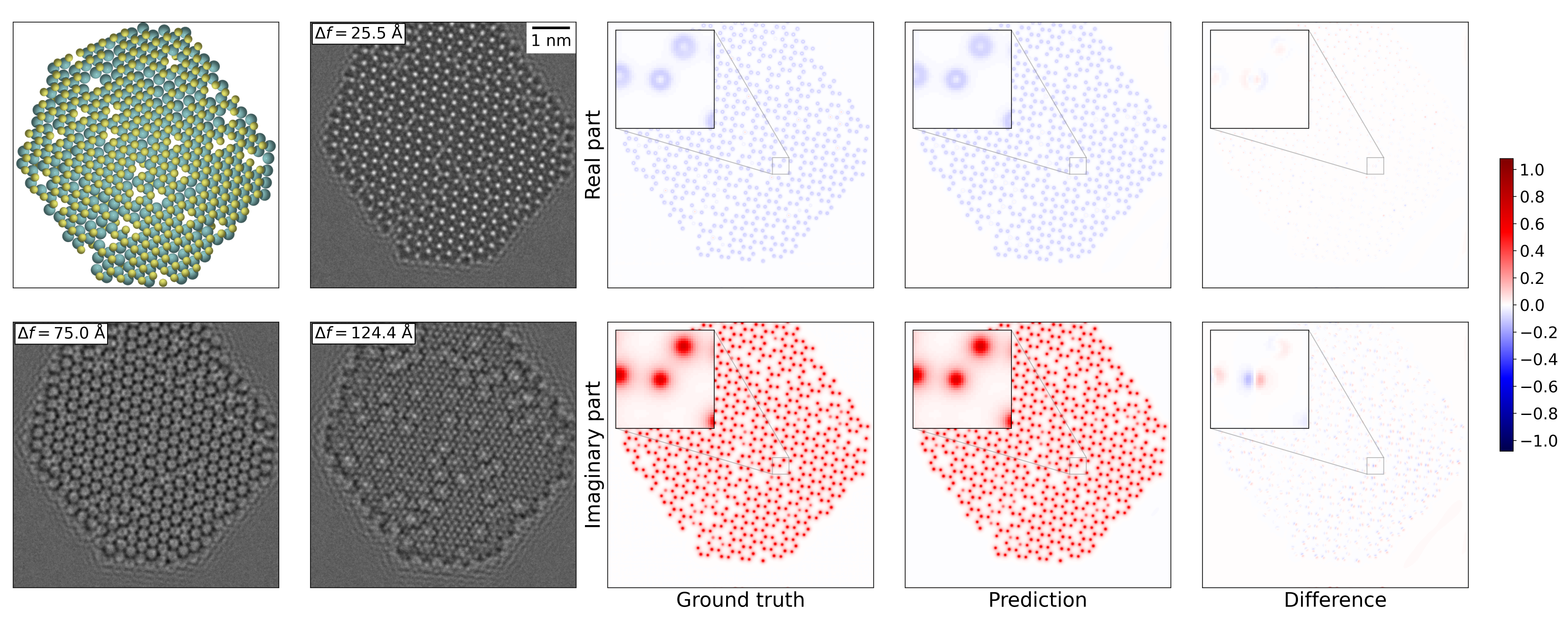}
  \caption{Reconstruction of the exit wave of a MoS$_2$ nanoparticle
    by a network trained on unsupported MoS$_2$ nanoparticles.
    The first two columns show the structure and the three simulated
    HRTEM images.  The third column shows the real and imaginary part
    of the actual wave function (the ground truth).  The fourth column
    shows the exit wave reconstructed by the neural network, and the
    last column shows the difference.
    The inset 
    highlights the point with the largest deviation, an atom misplaced
    by 7 pm (0.6 pixels).  The Root Mean
    Square Error (RMSE) is 0.0062.  The colormap
    indicates the scale used for plotting the exit wave, the same
    scale is used in all the following figures.}
  \label{fig:MoS2unsup}
\end{figure*}

Figure \ref{fig:MoS2unsup} shows the simplest situation, where the
network is trained and tested with unsupported MoS$_2$ nanoparticles.
The figure shows the real and imaginary parts of the exit wave used to
simulate the images (the ``ground truth''), and the exit wave
reconstructed by the neural network (the ``prediction'').  For thin
samples, the interaction between the electron wave and the sample
mainly results in a phase shift of the wave \cite{Chen2021exitwave},
this is also the case for the data in the figure, where the main part
of the signal is in the imaginary part.

The difference plot in Fig.~\ref{fig:MoS2unsup} shows that the
network clearly reconstructs the imaginary part of the exit wave
both qualitatively and quantitatively.  We see that all peaks are
reconstructed correctly, and that the neural network both reconstructs
the periodic lattice and the deviations from periodicity such as
vacancies, including single sulphur vacancies where a single sulphur
atom leaves a weaker peak than the usual two atoms. The system shown
in Figure~\ref{fig:MoS2unsup} was chosen as the median of the
validation set, half the systems in the validation set perform worse,
and half perform better.  In the SOI Section
S5 we show some of the worst systems in the
validation set, even the five percentile sample is reconstructed quite
well.

\begin{figure*}[t]
  \centering
  \includegraphics[width=\linewidth]{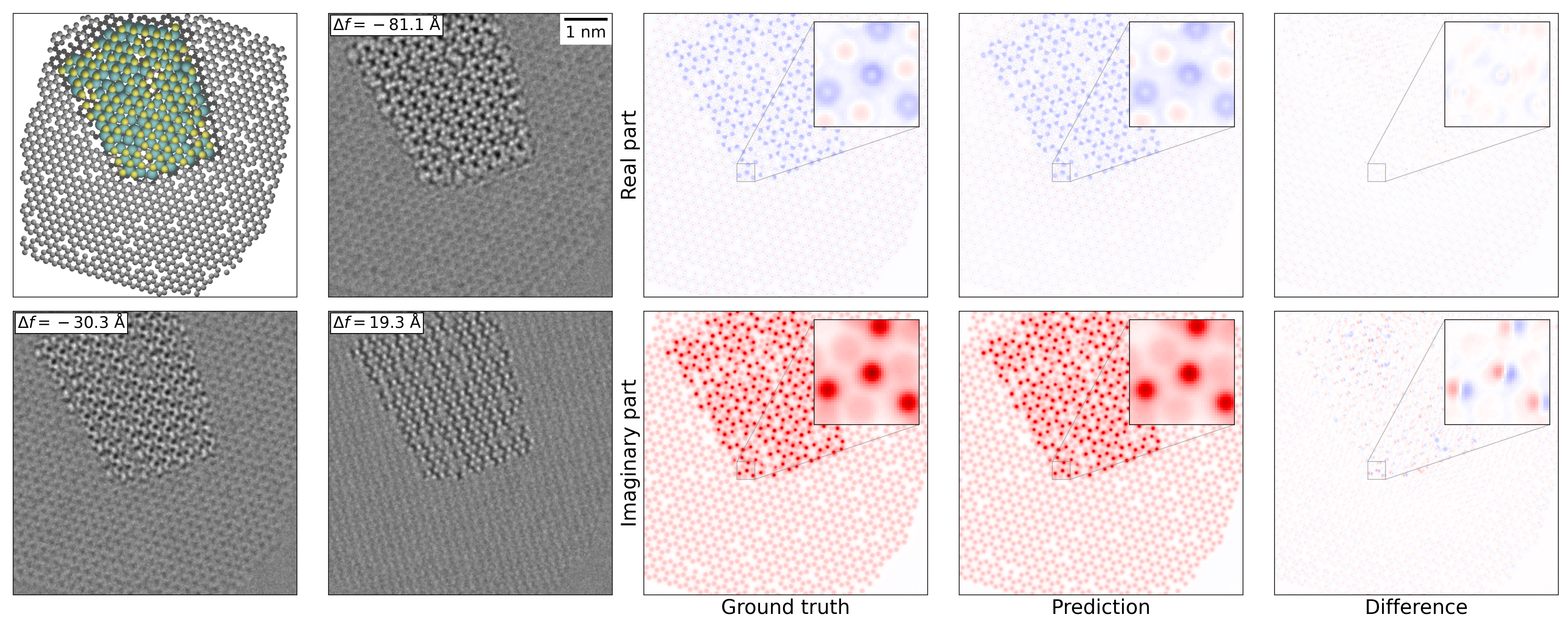}
  \caption{Reconstruction of the exit wave of a MoS$_2$ nanoparticle
    supported on graphene, by a network trained on graphene-supported
    MoS$_2$ nanoparticles.  The panels are the same as in
    Fig.~\protect\ref{fig:MoS2unsup}.  It is seen that the network
    locates the atoms both in the MoS$_2$ nanoparticle, and in the
    substrate.  The worst spot in the prediction where an atom is
    misplaced by a single pixel.  It is worth noticing that the
    graphene support is also reconstructed correctly, including the
    vacancies in the graphene.  The RMSE is 0.0122 and the colorbar is
    the same as in Fig.~\ref{fig:MoS2unsup}}
  \label{fig:MoS2sup}
\end{figure*}

Figure \ref{fig:MoS2sup} shows the more complex situation,
where the network is trained on graphene-supported 
MoS$_2$ nanoparticles.  The way the training set is constructed does
not guarantee that the full MoS$_2$ nanoparticle is overlapping with
the support, so in this case the network needs to learn to recognize
both supported and unsupported MoS$_2$.

The
network is able to reconstruct both the part of the wave function
coming from the support and from the nanoparticle, in spite of the
signal from the support being much weaker than from the nanoparticle.  The
network is even able to correctly find the carbon vacancies that have
been introduced in the support.   It should be noted that if the network
is trained for a shorter time (50 epochs instead of 200), it loses its
ability to find the carbon atoms below the nanoparticle.
The largest deviation in the reconstructed exit wave comes from a
slight misplacement of the atoms in the MoS$_2$ layer, the maximal
error in the placement of an atom is 9.7 pm, corresponding to a single
pixel.   This system is again chosen as the median of the validation
set.  

\begin{figure*}
\centering
\includegraphics[width=\linewidth]{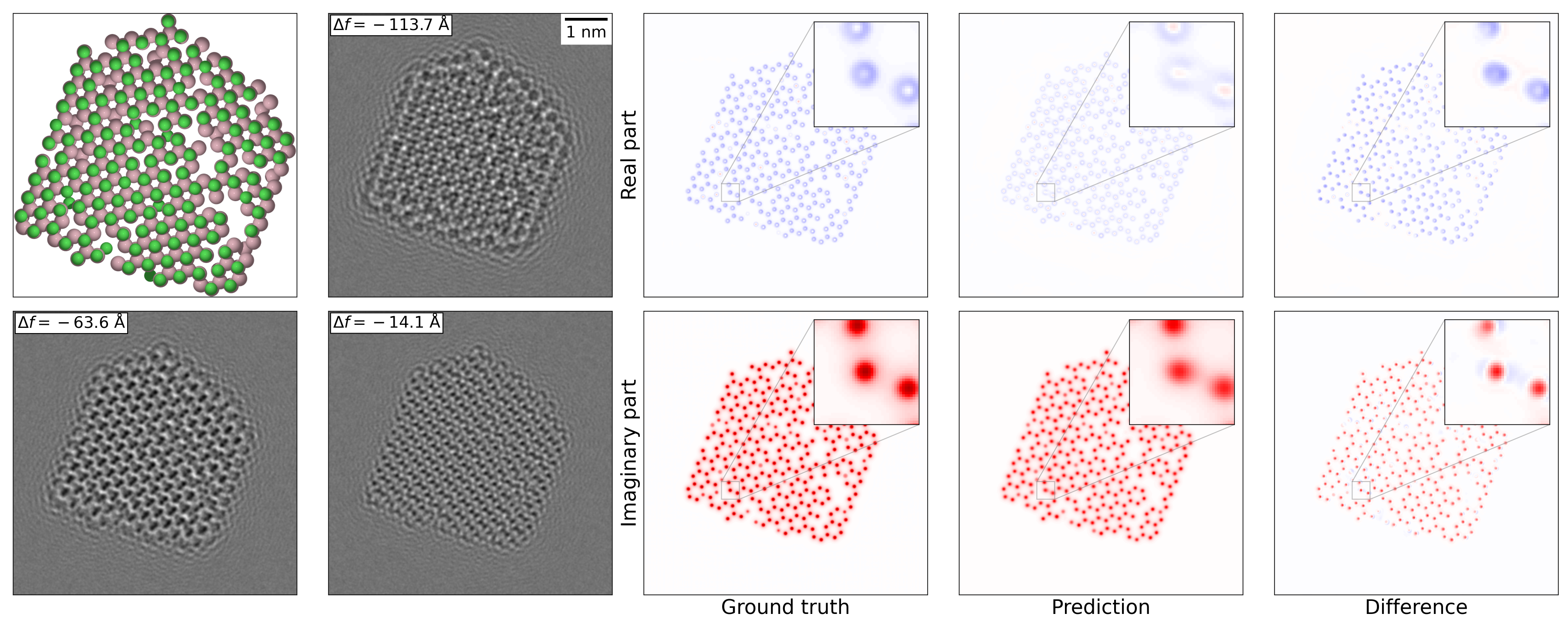}
 \caption{Reconstruction of the exit wave of a CoCl nanoparticle
   by a network trained on the C2DB (see text).  The panels are the same as in
   Fig.~\protect\ref{fig:MoS2unsup}.  The network correctly determines the
   positions of all the atoms (the maximal deviation is 8 pm or 0.77 pixel)), but does not correctly reproduce the
   sharpnes of the peaks in the wave function.  The RMSE is 0.0263.}
 \label{fig:c2db}
\end{figure*}

Finally, the method was tested on the C2DB database of 3393 proposed
two-dimensional materials \cite{Haastrup:2018ca}.  Again we show the
median system, a nanoparticle of CoCl, (Fig.~\ref{fig:c2db}).  We see
how all atoms are placed correctly, but the detailed shape of the
peaks in the imaginary part of the wave function is not well
reproduced, the network predicts somewhat smoother peaks.  In
addition, the network does not always identify positions where single
atoms are missing, leaving only one atom in the atomic column.  Each
position in the apparent hexagonal lattice contain both a Co and a Cl
atom, alternately oriented with the Co or Cl on top, and staggered in
the $z$ direction.

\begin{figure}
  \centering
  \includegraphics[width=\linewidth]{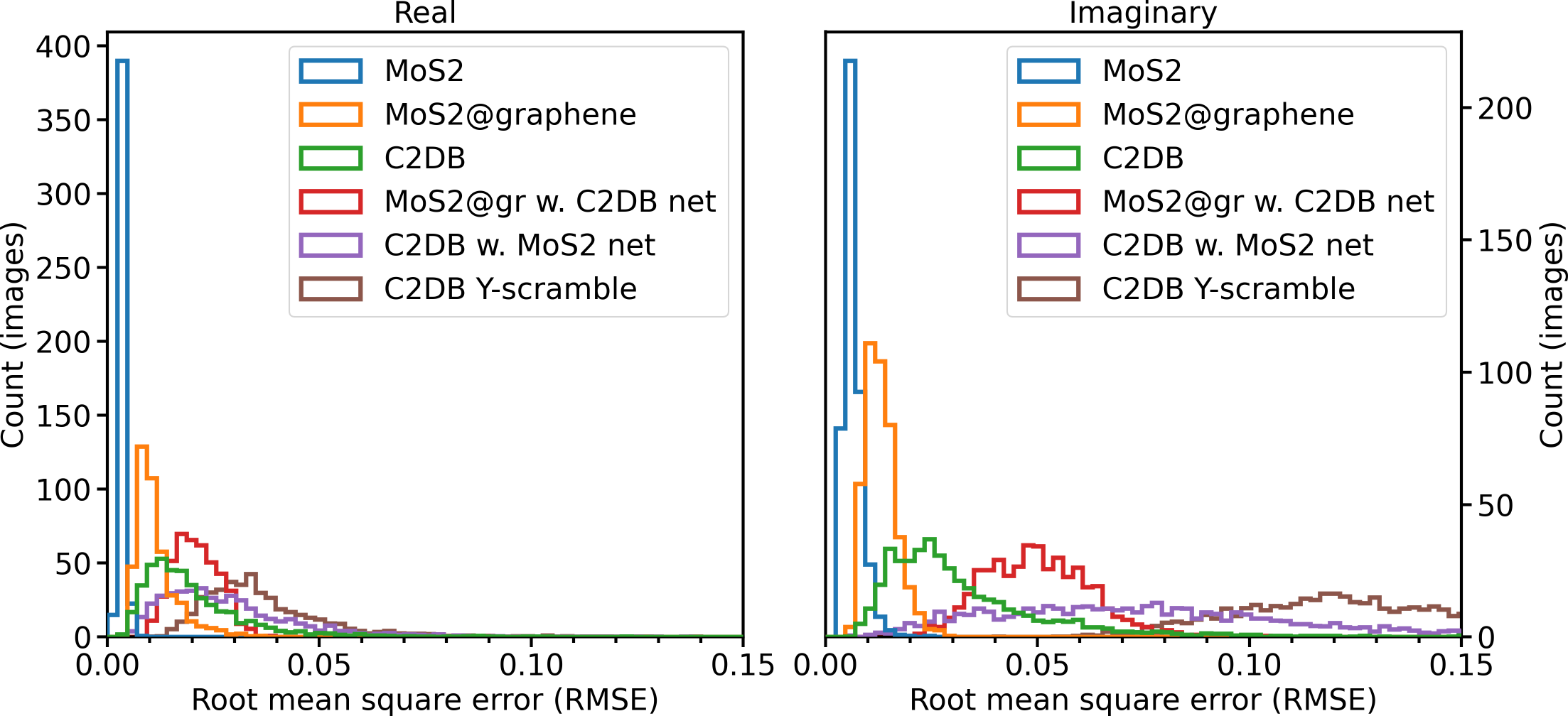}
  \caption{Histograms of the root-mean-square error of the images in
    the validation sets for the various networks, showing their
    relative performance.   Blue is the network
    trained and tested on unsupported MoS$_2$, orange is graphene supported MoS$_2$, and green is the
    C2DB.  It is seen that the performance of the network decreases
    somewhat as the complexity of the data set is increased.  The
    brown line is a baseline, this is the performance
    obtained if the network does not at all recognize the structure,
    obtained using the Y-scramble method (see text).    The red curve shows the
    supported MoS$_2$ validation set with the network trained on the more diverse
    C2DB.  As the samples contain two separate lattices, it is outside the training set of the C2DB.   
    Validating the C2DB test set with the network trained on MoS$_2$
    also gives bad results (purple curve), as the C2DB contains structures
    too far from what is observed in MoS$_2$.  }
  \label{fig:histotypes}
\end{figure}

In order to obtain a more quantitative measure of the performance of
the networks, we have created histograms of the root-mean-square error
(RMSE) of all the images in the validation sets, see
Fig.~\ref{fig:histotypes}. In general, the networks is better at
reproducing the strong signal in the imaginary part of the exit wave
than the weaker real part.  It is seen that the performance of the
network decreases somewhat as the complexity of the data set is
increased, going from unsupported MoS$_2$ to supported MoS$_2$ to the
C2DB dataset.  It is not surprising that the network can be trained
for better performance on the simpler datasets.  As a ``baseline'', we
also show the histogram produced from one of the datasets where the
predictions are compared with randomly chosen other exit waves in the
dataset (the Y-scramble method) rather than with the correct exit
wave.  This shows the performance of a hypothetical network learning
the overall properties of exit waves but learning nothing about the
specific systems, i.e.\ it acts as a ``null hypothesis''.

\begin{figure}[t]
  \centering
  \includegraphics[width=\linewidth]{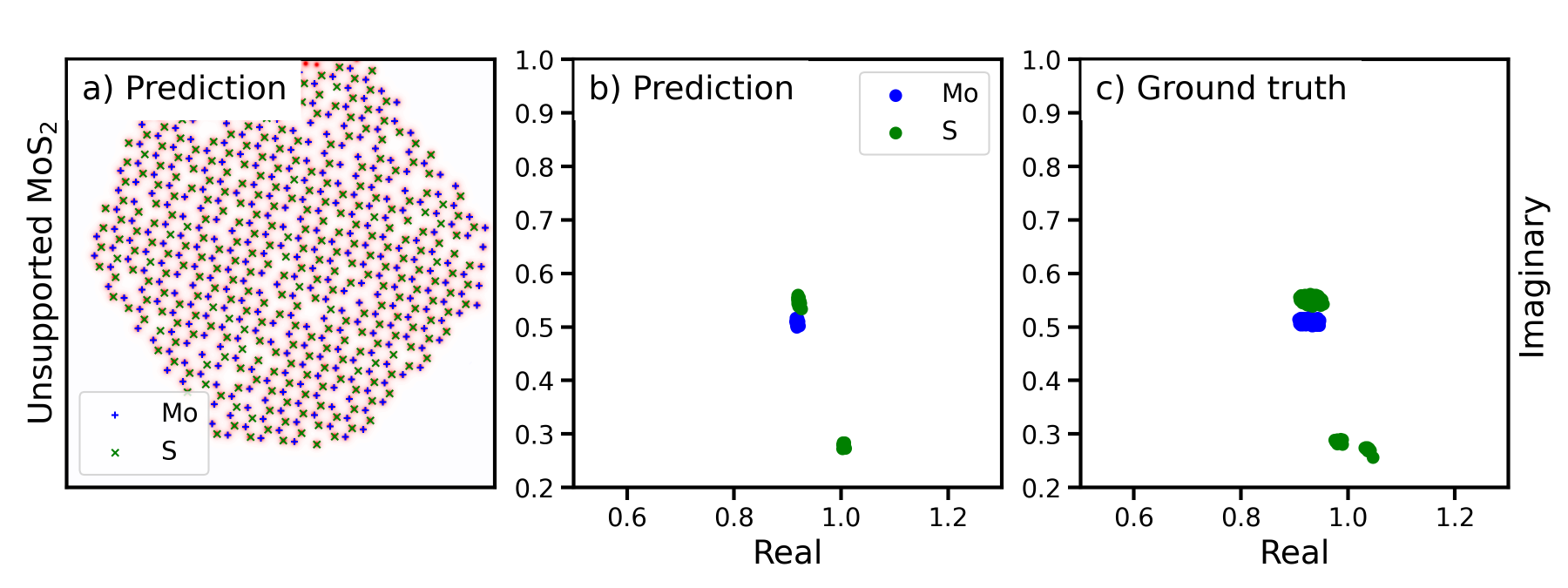}
  \includegraphics[width=\linewidth]{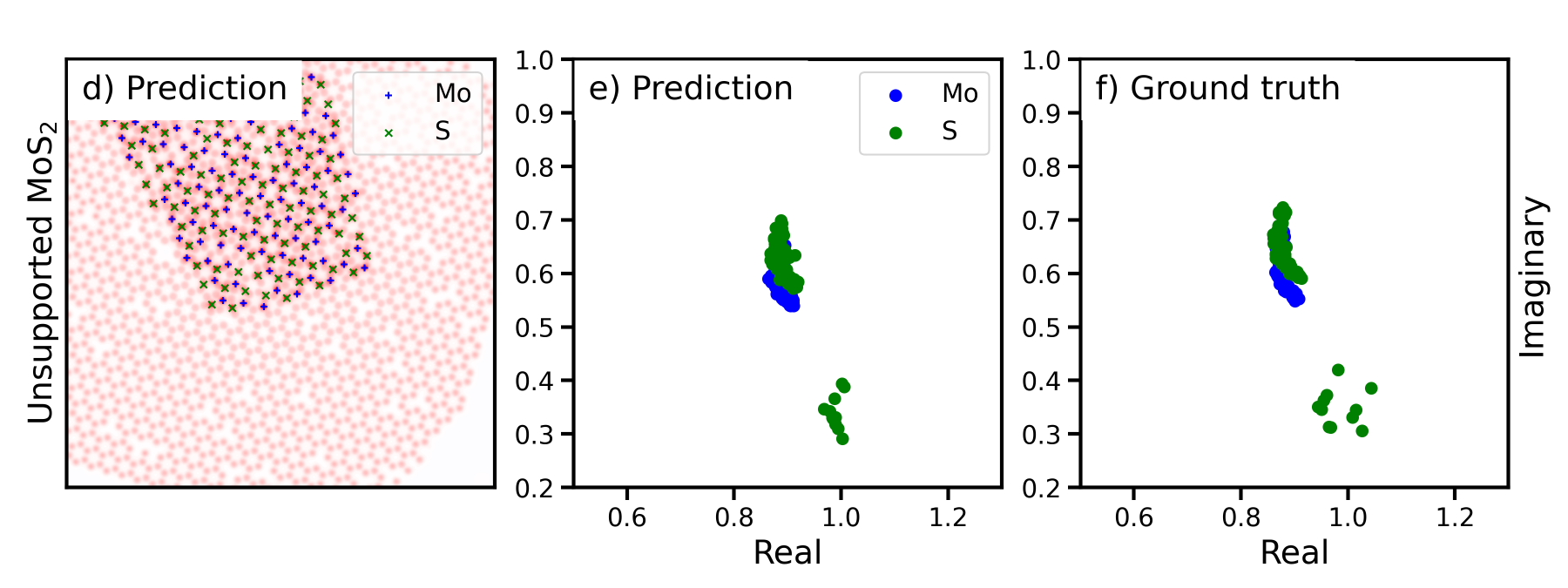}
  \caption{Argand plots of the complex value of the exit wave function
    at the local maxima of the change in wave function
    ($\left| \Psi_{\text{exit}} - 1 \right|$).  Top row: unsupported
    MoS$_2$ (the same system as Fig.~\protect\ref{fig:MoS2unsup}).
    Bottom row: supported MoS$_2$ (same as
    Fig.~\protect\ref{fig:MoS2sup}).  a+d: The imaginary part of the
    reconstructed wave function.  The peaks are marked with blue
    plusses or green crosses, depending on whether they correspond to
    a Mo or S atomic column position.  b+e: The Argand plots of the
    reconstructed wave function.  The separate points in the lower
    part of the plot corresponds columns with a single S atom instead
    of two.  c+f) The similar Argand plots made from the ground truth
    exit wave function.}
  \label{fig:argand}
\end{figure}

It is also seen that the relative error is significantly larger for
the real part of the exit wave. This is because its magnitude is 3--4
times smaller than the imaginary part (this can e.g.\ be seen by the
position of the peaks in the Argand plots in Figure \ref{fig:argand}).
It is only in the simplest case (unsupported MoS$_2$) that the network
performs well on the real part.

We also test how networks trained on the C2DB dataset performs on
the supported MoS$_2$ and vice versa.  Unsurprisingly, the network
trained on supported MoS$_2$ performs poorly on the C2DB dataset, as
the latter contains a far richer variety of structures.  On the other hand,
the network trained on the C2DB generates a very broad distribution of
results when applied to the database of supported MoS$_2$ structures
(the red curve in Fig.~\ref{fig:histotypes}).  Our interpretation is that this is because the
network correctly analyses the parts of the system where the MoS$_2$
and graphene only overlap a little, but performs badly where they
overlap.  While these systems
are not inherently more complicated than in the C2DB, they
differ in a fundamental way, as there are two different lattices in
the system (the lattice of graphene and the one of MoS$_2$), whereas
all systems in the C2DB training set only contain a single (but often
more complicated) crystal lattice.  This illustrates the importance of
training the network on systems that are similar to the final
application.

The purpose of an exit wave reconstruction is usually to extract
quantitative information about the atomic columns.  This is often done
in form of an Argand plot, where the peak values of the wave function at
the locations of the atoms are plotted in the complex plane
\cite{Chen:2015ik}.  It is therefore not enough that an exit wave
reconstructed by neural networks visually and statistically ressemble
the actual exit wave, it should also permit analysis in an Argand
plot.  This is shown in Fig.~\ref{fig:argand}, where we show Argand
plots of both the unsupported and supported nanoparticles from
Figs.~\ref{fig:MoS2unsup} and \ref{fig:MoS2sup}.  For the unsupported
nanoparticle, the Argand plot is just able to distinguish between a
single Mo atom (atomic number $Z=42$) and a sulphur dimer (sum of
atomic numbers $\sum Z = 32$).  The sulphur vacancies, where there is
only a single sulphur atom in the atomic column ($Z = 16$) are clearly
separated from the other types.  It is, however, not possible to
determine if the missing atom was above or below the plane of the Mo
atoms, although that information was present in the original wave
function (shown as ``ground truth'', where we see that the spots
corresponding to single S atoms is split into two nearby spots with
different real value, as would be expected from atoms with different
$z$ coordinate, see Chen \emph{et al.} \cite{Chen:2015ik}).

For the case of supported MoS$_2$ [Fig. \ref{fig:argand}(d-f)], the
picture is less clear.  The Argand plot still clearly separates the
sulphur vacancies from the other atomic columns, but there is a larger
spread on the column values, and no
longer a clear separation between columns containing two S atoms or a
single Mo atom.
However, if the same analysis is done on the ground truth exit waves
(Fig~\ref{fig:argand}(f)), the situation is the same.  This is most
likely due to interference from the substrate.

In the Argand plot, the position along the imaginary axis is largely
indicative of the total atomic number of the atomic column in the weak
phase limit \cite{Chen2021exitwave}.  The
positions of the Argand points are also affected by the $z$-height of the
column relative to the plane of the exit wave.  This is mainly due to
the peak spreading out spatially as the wave propagates from the
bottom of the atomic column to the plane where the exit wave is
defined, leading to a decrease in the peak value for atomic columns at
higher $z$ \cite{vanDyck2012bigbang}.  This effect is clearly not reproduced by the neural
network, as it cannot distinguish between single sulphur vacancies on
the two sides of the nanoparticle [Fig.~\ref{fig:argand}(b+c)].  It is
possible that a network could be trained to distinguish these features by
including training data where they are more prominent, i.e.\ a larger
concentration of
single sulphur vacancies and perhaps samples with higher tilt angles,
producing height differences.

\begin{figure}
  \centering
  \includegraphics[width=\linewidth]{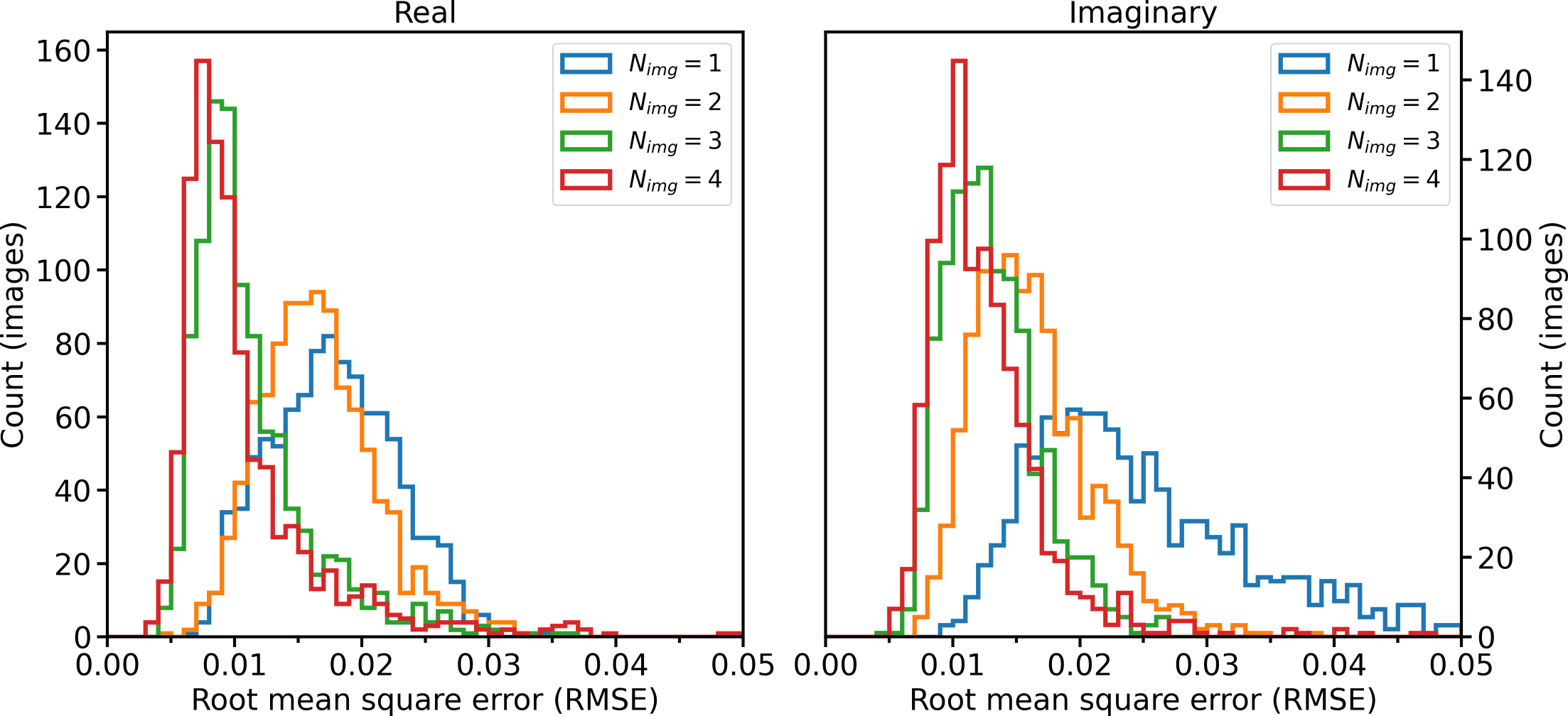}
  \caption{Test of how the number of input images affects the
    performance.  A single input image (blue curve) clearly does not
    give a good reconstruction of the exit wave. Already with two images (orange
    curve), good performance is obtained, at least for the imaginary
    part. Three images
    (green) as used in the rest of this work gives an
    improvement, whereas four images (red) gives only a marginal further
    improvement.}
  \label{fig:numimages}
\end{figure}

As a significant amount of information about the exit wave is encoded
in how the image changes with defocus, it is our working hypothesis
that a number of images are necessary for a neural network to be able
to reconstruct the exit wave.  This is verified in
Fig.~\ref{fig:numimages}, showing the performance of networks trained
on the same C2DB training sets but with a different number of input
images.  It is seen that some information about the exit wave can be
gained from even a single image, but a dramatic improvement is seen
going to two input images.  A small further improvement is seen when
increasing the number of images to three or four, and we decided to
use three images in the rest of this work.  In the simulations with two, three or four images, the
total range of defocus from the first to the last image were in each
case 10 nm.

\section{When the network fails}
\label{sec:when-network-fails}

No neural network is perfect, and it is important to be aware of the
kind of failures that can occur when analysing an image.  We
illustrate this with two kinds of errors observed in the C2DB
database.

\begin{figure}
\centering
\includegraphics[width=\linewidth]{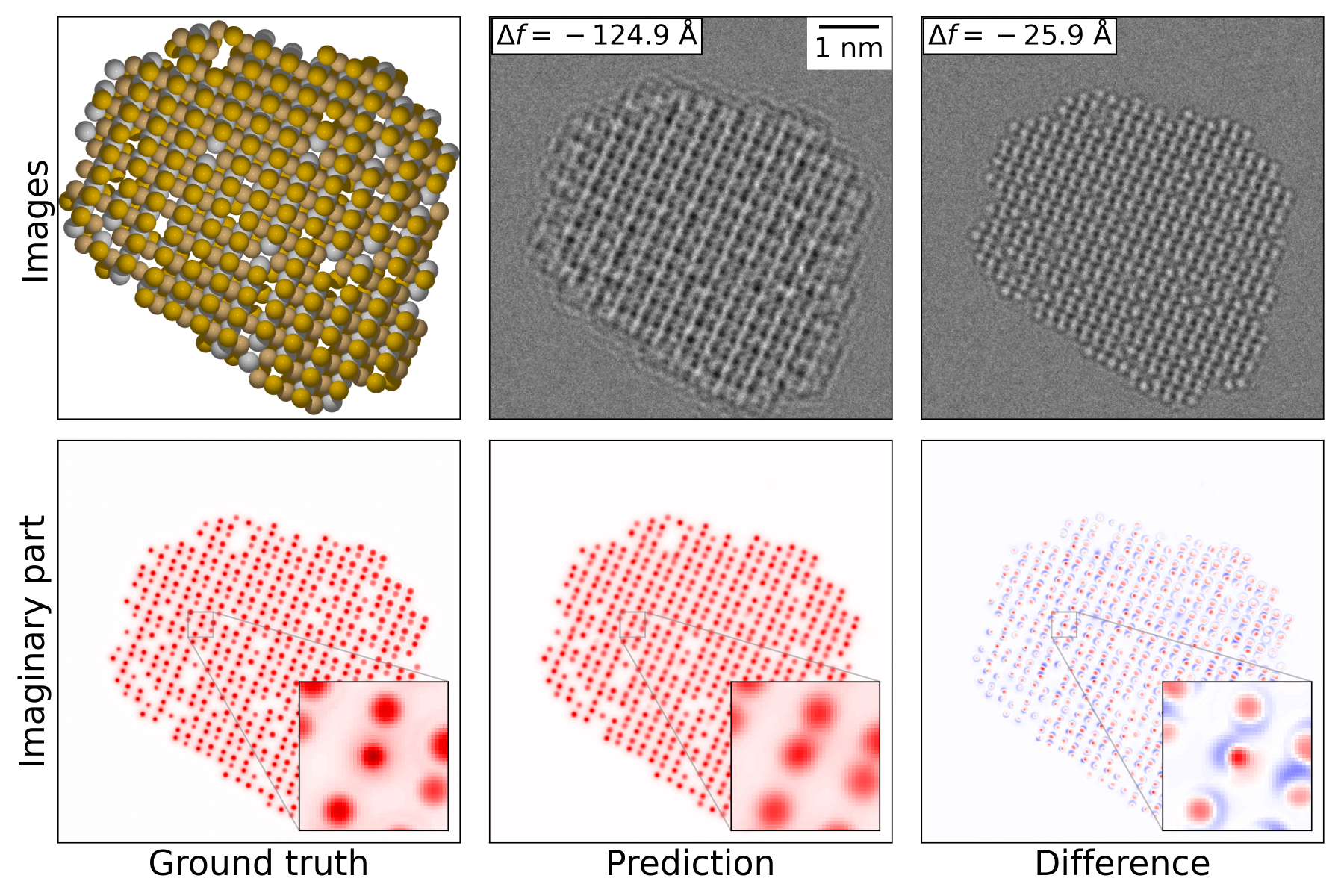}
 \caption{Reconstruction of the exit wave of a AgCuTe$_2$ nanoparticle
   by a network trained on the C2DB.  To save space we do not show the
   real part of the exit wave, and only the first and the last of the
   three images.  The network does not recognize that the
   copper atoms are systematically slightly offset from the high-symmetry
   positions.}
 \label{fig:brokensymmetry}
\end{figure}

The first case is silver copper telluride (AgCuTe$_2$), shown in
Fig.~\ref{fig:brokensymmetry}.  On one hand, the method
reliably finds all the vacancies in the structure, a task that would
be very difficult by visual inspection of the three images.  On
the other hand, the network fails to discover a small spontaneous breaking
of the symmetry in the structure: the Cu atoms are slightly displaced
compared to the rectangular lattice formed by the Ag and Te atoms. This is a highly
unusual configuration, and the neural network interprets it as the far
more common symmetric configuration. 

\begin{figure}
  \centering
  \includegraphics[width=\linewidth]{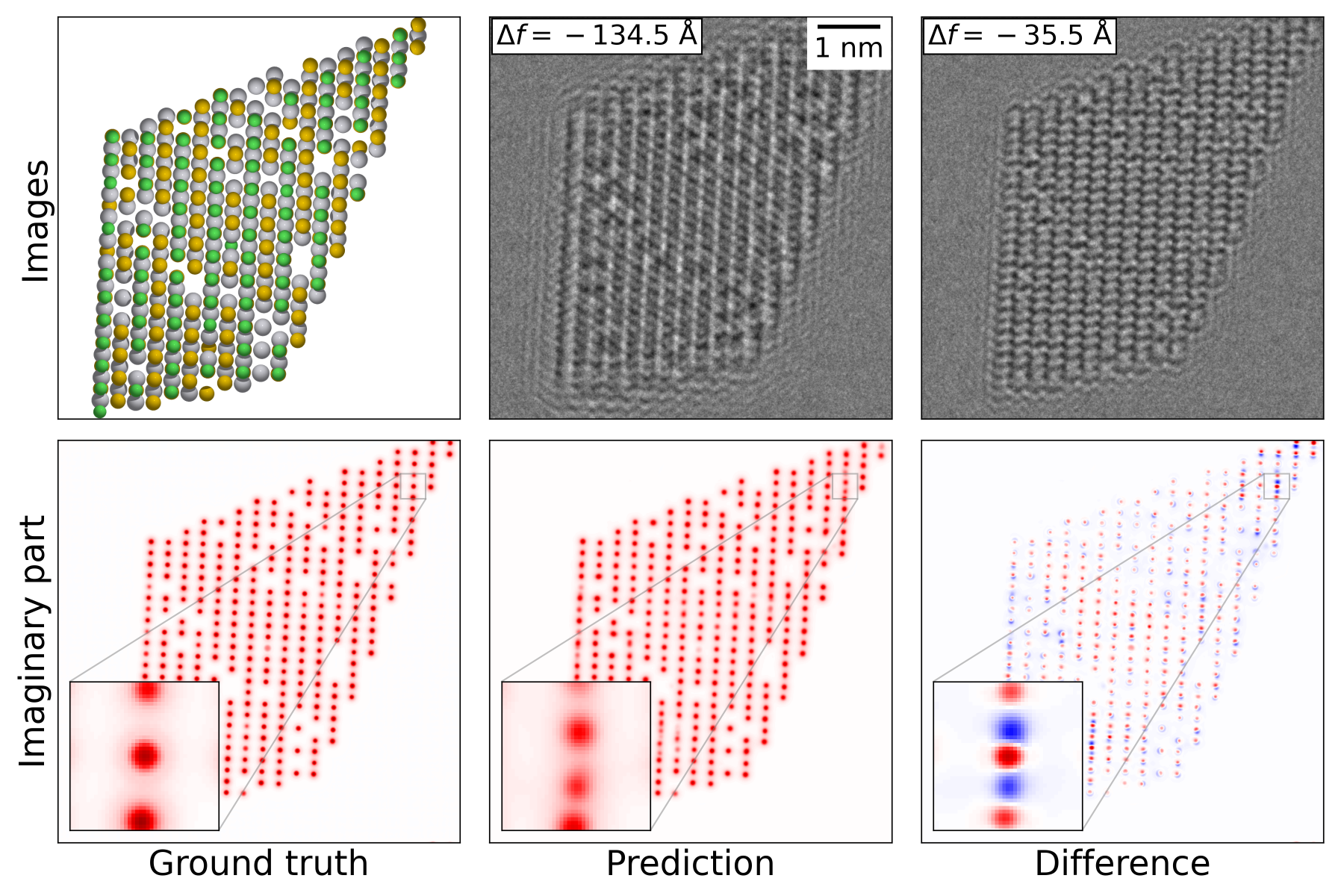}
  \caption{An example of the neural network inserting extra atoms in
    several places in the system.  The system is PtSeCl.}
  \label{fig:extraatoms}
\end{figure}

In the second case, the network is locally inserting extra atoms into
the structure, creating weird unphysical defects, see
Fig.~\ref{fig:extraatoms}.  This kind of errors should be relatively
easy to spot for the scientist.

The cases in Figs.~\ref{fig:brokensymmetry} and \ref{fig:extraatoms}
were chosen manually.  In the SOI, we give examples of some of the
worst and best results of the networks, selected solely from the RMSE
of the prediction.

As the examples here show, it is difficult to train a single network
to 3400 different materials, even if they are two-dimensional.  The
networks trained to a single material (MoS$_2$), with or without
support, do not exhibit these failure modes.  It is therefore
recommended to train networks to smaller classes of materials matching
the kinds of systems being studied experimentally.  Furhermore, the
kinds of errors shown here can be detected by training two or more
different networks to similar data sets, and detecting when the
networks differ in their prediction.

\section{Application to experimental data}
\label{sec:experimental}

We apply the method to experimental data, a focal series of a
MoS$_2$ model catalyst recorded on the TEAM 0.5 transmission electron
microscope at 50 keV beam energy.  The data analysed here is similar
to what was published recently by Chen \emph{et al.}\
\cite{Chen2021exitwave}, and we refer to that publications for details
regarding the experimental setup.

In their publication, Chen \emph{et al.}\ used 
focal series of 20 -- 44 images to reconstruct the wave functions.  Here, we
have selected three images from their focal series for analysis by the
neural network.

\begin{figure}
  \centering
   \includegraphics[width=\linewidth]{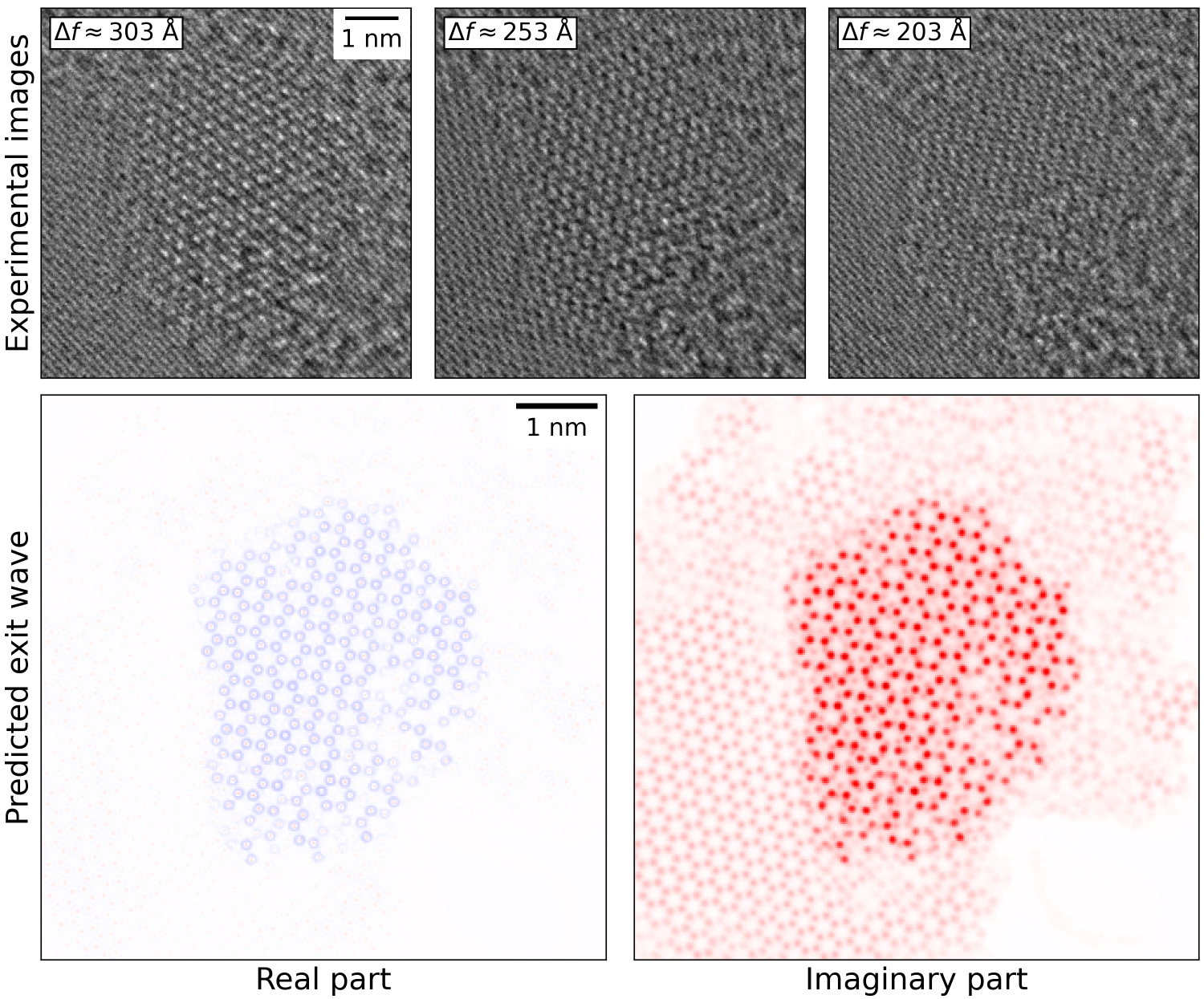}
   \caption{Applying the neural network to three experimental images
     (top row) results in the exit wave function depicted in the
     bottom row.  The network is clearly able to identify atomic
     positions in the MoS$_2$ nanoflake, but is not able to
     distinguish between Mo and S$_2$.  The network also provides a
     best guess on the positions of carbon atoms in the support, but
     as the support is graphite and the network was trained on single
     graphene layers as support, that cannot be considered reliable.
     The defocus values are as reported by MacTempas (overfocus),
     which uses the opposite sign convention from abTEM and this work.
     The colors are the same as in Fig.~\ref{fig:MoS2unsup}.}
  \label{fig:experimental}
\end{figure}

As the resolution of this image series is significantly lower than
what we have otherwise been using in this work (0.227 \AA/pixel
instead of 0.105 \AA/pixel) we retrained a network for this
resolution, based on the same data set of supported 
MoS$_2$, but resampled to resolutions in the interval from 0.215 to
0.235 \AA/pixel.  The lower resolution
had only a small detrimental effect on the network performance when
tested on the validation set.
We then selected three experimental images with a difference in
defocus of 50 \AA, to match the defocus difference between the three
images used to train the network.
The images and the
resulting exit wave are shown in Fig.~\ref{fig:experimental}.  As can
be seen, a clear exit wave is reconstructed, showing the honeycomb
lattice of the supported MoS$_2$ nanoflake, and of the supporting
graphite lattice.  However, an Argand plot is not able to 
distinguish the lattice points of the Mo and S sublattice (not shown),
consistent with what we saw in simulated images (see Figure
\ref{fig:argand}, panel e and f).  In both cases, the reason is the
same.  Some peaks in the wave function of the MoS$_2$ coincide with
peaks from the graphite, some do not, and that leads to greater
variation between peaks than the difference between a Mo atom and two
S atoms.

In their publication, Chen \emph{et al.} \cite{Chen2021exitwave} were
able to distinguish between peaks from Mo and S atomic columns, but
their analysis of the exit wave is also more elaborate.  First, they
had to Fourier filter their images, removing spatial frequencies
coming from the graphene support from the exit wave. Second, even if a
clear distinction of the peak imaginary values of the Mo and two S
atomic columns were made, it is worth noticing that the chemical
interpretation of the relative intensities calls for caution. As
reported by Chen \emph{et al.}, the peak values can be severely
reduced and the imaginary parts be broadended across a nanocrystal due
to heterogeneous vibrations response of the sample under
illumination. Chen \emph{et al.} offers a framework for an
interpretation of the exit wave function. This interpretation is
independent of the way in which the exit wave function is
reconstructed, which is the prime objective for the present analysis.

With even just a few images, the network can thus already capture the
main arrangement of the atomic columns based on an experimental focal
series of low-dose HRTEM images. Further inclusion of images from the
focal series might help in better account for the column intensitities
and role of high order aberrations on the contrast blurring in the
expeirmental image.  For a full qualitative analysis of the
experimental data, networks would have to be trained to specifically
take into account a more realistic model for the vibrations of the
atoms, as well as the more complicated multilayer support in the
experimental data.  In addition, the network should be trained to
handle carbon contamination of the sample.

\section{Comparison to traditional exit wave reconstruction}
\label{sec:comp-traditional}

To be able to compare this method with more traditional methods for
exit wave reconstructions, we have applied the algorithm of Gerchberg
and Saxton \cite{Gerchberg:1972ws}, as implemented in MacTempas
version 2.4.50, to three simulated image series of graphene supported MoS$_2$.
The systems were selected according to how well they had been reconstructed
by the neural network, we chose the 25, the 50 and the 75
percentile images (Figures S10, \ref{fig:MoS2sup} and
S11, respectively).

The generated data sets contain eleven images with a 1 nm change in
defocus between each of them, leading to a total defocus range of 10
nm, the same that was used for the neural networks.  All eleven images
are used for the Gerchberg-Saxton (GS) exit wave reconstruction, whereas only
three (the first, middle and last) were used for reconstructions with
the neural network. 

The GS exit wave reconstruction algorithm was given the actual values of
the defocus, the 
spherical aberration ($C_s$) and the focal spread, instead of
determining them through an optimization process as is usually done.
No coma or 2-fold astigmatism was assumed in the reconstruction
process, although both coma and astigmatism were present in the images.

In contrast, the neural network does not require any of this
information, it is trained to reconstruct the wave function a few
Ångström below the lowest atom in the sample without further knowledge
of neither the exact values of the defocus, nor of the aberrations of
the microscope, except that they are within the intervals used to
train the neural network (Table \ref{tab:parameters}).

\begin{figure}
  \centering
   \includegraphics[width=\linewidth]{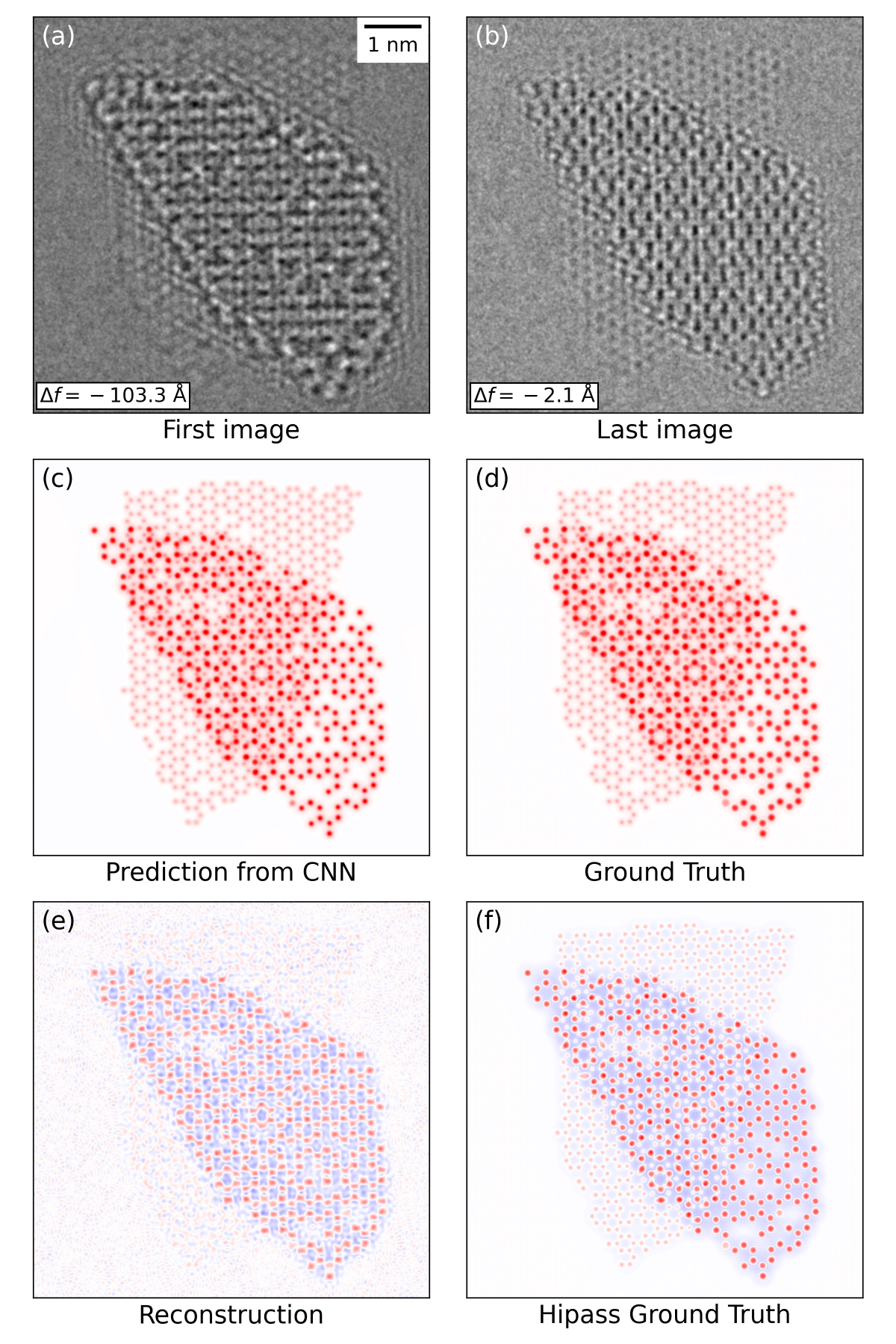}
   \caption{Comparing the neural network with a traditional algorithm
     for exit wave reconstruction. (a) and (b) The first and last
     images in the image series.  (c) The exit wave reconstructed by
     the network.  (d) The ground truth (correct) exit wave.  (e) The
     exit wave reconstructed by the Garchberg-Saxton algorithm.  The
     large deviations are due to the long wavelength part of the exit
     wave not being reconstructed.  (f) The ground truth wave function
     with the longest wavelengths removed.  The colormap for the exit
     wave is identical to the one in Fig.~\ref{fig:MoS2unsup}.}
  \label{fig:reconstcomp}
\end{figure}

A comparison between the neural network and the more traditional exit
wave reconstruction is shown in Fig.~\ref{fig:reconstcomp}.  At first
sight, it looks like the neural network strongly outperforms the
traditional reconstruction, the difference between the reconstructed
image and the original ground truth wave function is much smaller for
the neural network reconstruction.  However, this is mainly because the longest wavelengths in the exit wave that have not
been reconstructed by the Gerchberg-Saxton algorithm, 
leading to the phase of the wave function locally being averaged to
zero.  It is thus more fair to compare the reconstructed wave in
Fig.~\ref{fig:reconstcomp}(e) with a ground truth wave function where
the longest wavelengths have been filtered out (panel f), using a
Gaussian filter with a width of eight pixels (0.9 \AA).  In 
this case, visual inspection indicate that the error of the two 
models are of similar magnitude, although the neural network appears
to be performing best. This is confirmed by calculating the Root Mean
Square Error for the CNN reconstruction (i.e. for the difference
between panel c and d) and for the Gerchberg-Saxton (panels e and f).
The RMSE is 0.013 and 0.061, respectively.

The system shown in Figure \ref{fig:reconstcomp} is the 25-percentile
system.  Similar plots for the 50 and the 75-percentile systems are
shown in the SOI (Figures S13 and S14).  It should be noted that the
Gershberg-Saxton reconstruction of the 50-percentile image is of
significantly lower quality than the two others, although the neural
network did not have problems with this image series.  This could be
due to those images having both 2-fold astigmatism and coma in the
upper end of the range shown in Table \ref{tab:parameters}.

Inclusion of more abberrations than the ones in Table
\ref{tab:parameters} might change these conclusions, and might require using more
images for the neural network reconstruction to be reliable.  It does,
however, appear that a neural network is able to quickly give a
reconstructed exit wave of a quality at least comparable to a traditional exit
wave reconstruction from only a few images.

\section{Conclusions}
\label{sec:conclusions}

Convolutional Neural Networks are a promising alternative to
traditional exit wave reconstruction, with the obvious advantage that
they only require a few images instead of a long image sequence, that
the data processing is fast enough to be done in real time at the
microscope, and that detailed knowledge of the aberration parameters
of the microscope is not needed.  It does, however, require that the
networks are optimized for the systems at hand.

As expected, the method works best for simpler systems, illustrated
here with unsupported and graphene-supported MoS$_2$ nanoparticles,
where the exit waves are reproduced with a fidelity that allows for
both qualitative and quantitative analysis.  For significantly more
complicated structures, illustrated here with the relatively diverse
C2DB dataset, the network overall performs well, but fails to
reconstruct some details in some of the more complex
materials. Nevertheless, even in the more complicated materials, the
majority of the structure including the positions of point defects is
recovered by the neural network.

One could hope that the neural network had learned to generally invert
the Contrast Transfer Function of the microscope.  That is, however,
not the case.  The network utilises knowledge about ``likely''
structures based on the kind of structures it has seen in the training
set, and must be trained on structures similar to the ones it will be
used to analyse.  On the other hand, this use of prior knowledge
of the systems is probably what enables the network to reconstruct the
exit wave based on only three input images, and without knowledge of
the actual parameters of the CTF.  It should be pointed out that
including further abberrations than the ones used in this work (Table
\ref{tab:parameters}) may require using more than three images as
input to the neural network.

In summary, we have demonstrated that neural networks can be trained
to reconstruct the exit wave function of a varied class of
two-dimensional materials, with only three HRTEM images with different
defocus as input to the network.  We can train and validate the network
on simulated data, and then apply it to analyse experimentally
obtained data, demonstrated here with the case of MoS$_2$ supported
on graphene.

\section*{Acknowledgements}

We would like to thank Sophie K. Kaptain and Dr.\ Daniel Kelly for
technical assistance in connetion with the MacTempas exit wave
reconstructions.

\section*{Funding}
The authors acknowledge financial support from the Independent
Research Fund Denmark (DFF-FTP) through grant no.~9041-00161B.
L.P.H. was financially supported by The Danish Council for Technology
and Innovation (08-044837). The Center for Visualizing Catalytic
Processes is sponsored by the Danish National Research Foundation
(DNRF146).

\section*{Availability of data and materials}
The code is available on github \cite{code}.  The trained networks, the
scripts and data used to train the network and generate all figures
except Fig.~\ref{fig:experimental} are available from the DTU Data
repository at doi:10.11583/DTU.15263655.  The experimental data used in
Fig.~\ref{fig:experimental} belong to the authors of
Ref.~\cite{Chen2021exitwave}.

\section*{Competing interests}
The authors declare that they have no competing interests.



\begin{thebibliography}{10}
\expandafter\ifx\csname urlstyle\endcsname\relax
  \providecommand{\doi}[1]{doi:\discretionary{}{}{}#1}\else
  \providecommand{\doi}{doi:\discretionary{}{}{}\begingroup
  \urlstyle{rm}\Url}\fi

\bibitem{Kalinin:2019ds}
S.~V. Kalinin, et~al.
\newblock {Lab on a beam—Big data and artificial intelligence in scanning
  transmission electron microscopy}.
\newblock MRS Bull. 44 (2019) 565 -- 575.
\newblock \doi{10.1557/mrs.2019.159}.

\bibitem{Spurgeon:2020jr}
S.~R. Spurgeon, et~al.
\newblock {Towards data-driven next-generation transmission electron
  microscopy}.
\newblock Nature materials  (2020) 1 -- 6.
\newblock \doi{10.1038/s41563-020-00833-z}.

\bibitem{Quan2016}
T.~M. Quan, D.~G.~C. Hildebrand, W.-K. Jeong.
\newblock {FusionNet: A deep fully residual convolutional neural network for
  image segmentation in connectomics}.
\newblock arXiv.org  (2016) 1612.05360.

\bibitem{Holm:2020fz}
E.~A. Holm, et~al.
\newblock {Overview: Computer Vision and Machine Learning for Microstructural
  Characterization and Analysis}.
\newblock Metall. Mater. Trans. A 37 (2020-09) 1 -- 15.
\newblock \doi{10.1007/s11661-020-06008-4}.

\bibitem{Azimi:2018cy}
S.~M. Azimi, D.~Britz, M.~Engstler, M.~Fritz, F.~Mücklich.
\newblock {Advanced Steel Microstructural Classification by Deep Learning
  Methods}.
\newblock Sci. Rep. 8 (2018-02) 2128 -- 14.
\newblock \doi{10.1038/s41598-018-20037-5}.

\bibitem{DeCost:2019ix}
B.~L. DeCost, B.~Lei, T.~Francis, E.~A. Holm.
\newblock {High Throughput Quantitative Metallography for Complex
  Microstructures Using Deep Learning: A Case Study in Ultrahigh Carbon Steel}.
\newblock Microsc. Microanal. 25 (2019-02) 21 -- 29.
\newblock \doi{10.1017/s1431927618015635}.

\bibitem{Lin2021:TEMImageNet}
R.~Lin, R.~Zhang, C.~Wang, X.-Q. Yang, H.~L. Xin.
\newblock {TEMImageNet training library and AtomSegNet deep-learning models for
  high-precision atom segmentation, localization, denoising, and deblurring of
  atomic-resolution images}.
\newblock Sci. Rep. 11 (2021) 5386.
\newblock \doi{10.1038/s41598-021-84499-w}.

\bibitem{Vincent2021:Denoise}
J.~L. Vincent, et~al.
\newblock {Developing and Evaluating Deep Neural Network-Based Denoising for
  Nanoparticle TEM Images with Ultra-Low Signal-to-Noise}.
\newblock Microsc. Microanal.  (2021) 1--17.
\newblock \doi{10.1017/s1431927621012678}.

\bibitem{Madsen:2018ey}
J.~Madsen, et~al.
\newblock {A Deep Learning Approach to Identify Local Structures in
  Atomic-Resolution Transmission Electron Microscopy Images}.
\newblock Adv. Theory Simul. 1 (2018) 1800037.
\newblock \doi{10.1002/adts.201800037}.

\bibitem{Ziatdinov:2017ct}
M.~Ziatdinov, et~al.
\newblock {Deep Learning of Atomically Resolved Scanning Transmission Electron
  Microscopy Images: Chemical Identification and Tracking Local
  Transformations.}
\newblock ACS Nano 11 (2017) 12742 -- 12752.
\newblock \doi{10.1021/acsnano.7b07504}.

\bibitem{Lee:2020bk}
C.-H. Lee, et~al.
\newblock {Deep Learning Enabled Strain Mapping of Single-Atom Defects in
  Two-Dimensional Transition Metal Dichalcogenides with Sub-Picometer
  Precision}.
\newblock Nano Lett. 20 (2020-05) 3369 -- 3377.
\newblock \doi{10.1021/acs.nanolett.0c00269}.

\bibitem{2020arXiv200110938E}
J.~M. Ede, J.~J.~P. Peters, J.~Sloan, R.~Beanland.
\newblock {Exit Wavefunction Reconstruction from Single Transmission Electron
  Micrographs with Deep Learning}.
\newblock arXiv.org  (2020) 2001.10938.

\bibitem{Meyer:2008je}
Meyer, Heindl.
\newblock {Reconstruction of off-axis electron holograms using a neural net}.
\newblock J. Microsc. 191 (2008) 52 -- 59.
\newblock \doi{10.1046/j.1365-2818.1998.00343.x}.

\bibitem{deBeeck:1996ii}
M.~O.~d. Beeck, D.~V. Dyck, W.~Coene.
\newblock {Wave function reconstruction in HRTEM: The parabola method}.
\newblock Ultramicroscopy 64 (1996-08) 167 -- 183.
\newblock \doi{10.1016/0304-3991(96)00058-7}.

\bibitem{Tiemeijer:2012em}
P.~C. Tiemeijer, M.~Bischoff, B.~Freitag, C.~Kisielowski.
\newblock {Using a monochromator to improve the resolution in TEM to below 0.5
  Å. Part II: application to focal series reconstruction.}
\newblock Ultramicroscopy 118 (2012-07) 35 -- 43.
\newblock \doi{10.1016/j.ultramic.2012.03.019}.

\bibitem{Chen:2015ik}
F.~R. Chen, C.~Kisielowski, D.~V. Dyck.
\newblock {3D reconstruction of nanocrystalline particles from a single
  projection.}
\newblock Micron 68 (2015-01) 59 -- 65.
\newblock \doi{10.1016/j.micron.2014.08.009}.

\bibitem{Chen2021exitwave}
F.-R. Chen, et~al.
\newblock {Probing atom dynamics of excited Co-Mo-S nanocrystals in 3D}.
\newblock Nature Comm. 12 (2021) 5007.
\newblock \doi{10.1038/s41467-021-24857-4}.

\bibitem{VanDyck:2015fg}
D.~V. Dyck, I.~Lobato, F.-R. Chen, C.~Kisielowski.
\newblock {Do you believe that atoms stay in place when you observe them in
  HREM?}
\newblock Micron 68 (2015) 158 -- 163.
\newblock \doi{10.1016/j.micron.2014.09.003}.

\bibitem{Thust1996exitwave}
A.~Thust, W.~Coene, M.~O.~d. Beeck, D.~V. Dyck.
\newblock {Focal-series reconstruction in HRTEM: simulation studies on
  non-periodic objects}.
\newblock Ultramicroscopy 64 (1996) 211--230.
\newblock \doi{10.1016/0304-3991(96)00011-3}.

\bibitem{Hsieh:2004ca}
W.-K. Hsieh, F.-R. Chen, J.-J. Kai, A.~I. Kirkland.
\newblock {Resolution extension and exit wave reconstruction in complex HREM.}
\newblock Ultramicroscopy 98 (2004-01) 99 -- 114.
\newblock \doi{10.1016/j.ultramic.2003.08.004}.

\bibitem{Allen:2004hc}
L.~J. Allen, W.~McBride, N.~L. O'Leary, M.~P. Oxley.
\newblock {Exit wave reconstruction at atomic resolution.}
\newblock Ultramicroscopy 100 (2004-07) 91 -- 104.
\newblock \doi{10.1016/j.ultramic.2004.01.012}.

\bibitem{Ferrari:2015co}
A.~C. Ferrari, et~al.
\newblock {Science and technology roadmap for graphene, related two-dimensional
  crystals, and hybrid systems}.
\newblock Nanoscale 7 (2015-03) 4598 -- 4810.
\newblock \doi{10.1039/c4nr01600a}.

\bibitem{Bhimanapati:2015bo}
G.~R. Bhimanapati, et~al.
\newblock {Recent Advances in Two-Dimensional Materials beyond Graphene.}
\newblock ACS Nano 9 (2015) 11509 -- 11539.
\newblock \doi{10.1021/acsnano.5b05556}.

\bibitem{Chorkendorff:2007book}
I.~Chorkendorff, J.~W. Niemantsverdriet.
\newblock {Concepts of Modern Catalysis}.
\newblock Wiley-VCH, Weinheim, 2 ed. (2007).

\bibitem{Ronneberger:2015gk}
O.~Ronneberger, P.~Fischer, T.~Brox.
\newblock {U-Net: Convolutional Networks for Biomedical Image Segmentation}.
\newblock vol. 9351 of \emph{Medical Image Computing and Computer-Assisted
  Intervention – MICCAI 2015}, pp. 234 -- 241. Springer International
  Publishing (2015).
\newblock \doi{10.1007/978-3-319-24574-4\_28}.

\bibitem{Chollet:2018uk}
F.~Chollet.
\newblock {Deep Learning with Python}.
\newblock Manning (2018).

\bibitem{tensorflow}
\url{https://www.tensorflow.org/}.

\bibitem{HjorthLarsen:2017hn}
A.~H. Larsen, et~al.
\newblock {The atomic simulation environment - a Python library for working
  with atoms.}
\newblock J. Phys.: Condens. Matter 29 (2017) 273002.
\newblock \doi{10.1088/1361-648x/aa680e}.

\bibitem{Haastrup:2018ca}
S.~Haastrup, et~al.
\newblock {The Computational 2D Materials Database: high-throughput modeling
  and discovery of atomically thin crystals}.
\newblock 2D Materials 5 (2018-10) 042002.
\newblock \doi{10.1088/2053-1583/aacfc1}.

\bibitem{Goodman:1974ku}
P.~Goodman, A.~F. Moodie.
\newblock {Numerical evaluations of N-beam wave functions in electron
  scattering by the multi-slice method}.
\newblock Acta Crystallogr. Sect. A 30 (1974-03) 280 -- 290.
\newblock \doi{10.1107/s056773947400057x}.

\bibitem{kirkland2010book}
E.~J. Kirkland.
\newblock {Advanced Computing in Electron Microscopy}  (2010).
\newblock \doi{10.1007/978-1-4419-6533-2}.

\bibitem{Madsen2021:abtem}
J.~Madsen, T.~Susi.
\newblock {The abTEM code: transmission electron microscopy from first
  principles}.
\newblock Open Research Europe 1 (2021) 24.
\newblock \doi{10.12688/openreseurope.13015.1}.

\bibitem{Mannebach:2015}
E.~M. Mannebach, et~al.
\newblock {Dynamic Structural Response and Deformations of Monolayer MoS2
  Visualized by Femtosecond Electron Diffraction}.
\newblock Nano Letters 15 (2015) 6889--6895.
\newblock \doi{10.1021/acs.nanolett.5b02805}.

\bibitem{kingma2017adam}
D.~P. Kingma, J.~Ba.
\newblock Adam: A method for stochastic optimization.
\newblock arXiv.org  (2017) 1412.6980.

\bibitem{vanDyck2012bigbang}
D.~V. Dyck, J.~R. Jinschek, F.-R. Chen.
\newblock {‘Big Bang’ tomography as a new route to atomic-resolution
  electron tomography}.
\newblock Nature 486 (2012) 243--246.
\newblock \doi{10.1038/nature11074}.

\bibitem{Gerchberg:1972ws}
R.~W. Gerchberg, W.~O. Saxton.
\newblock {A practical algorithm for the determination of phase from image and
  diffraction plane pictures}.
\newblock Optik 35 (1972) 237 -- 256.

\bibitem{code}
\url{https://gitlab.com/schiotz/NeuralNetwork_HRTEM/-/tree/ExitWave}.

\end{thebibliography}

\clearpage
\appendix
\renewcommand\thefigure{S\arabic{figure}}    
\section{Supplementary online information}
\label{sec:supplementary}
\setcounter{figure}{0}

\subsection{Network architecture}
\label{sec:network-architecture}

\begin{figure}[htbp]
  \centering
  \includegraphics[width=0.95\linewidth]{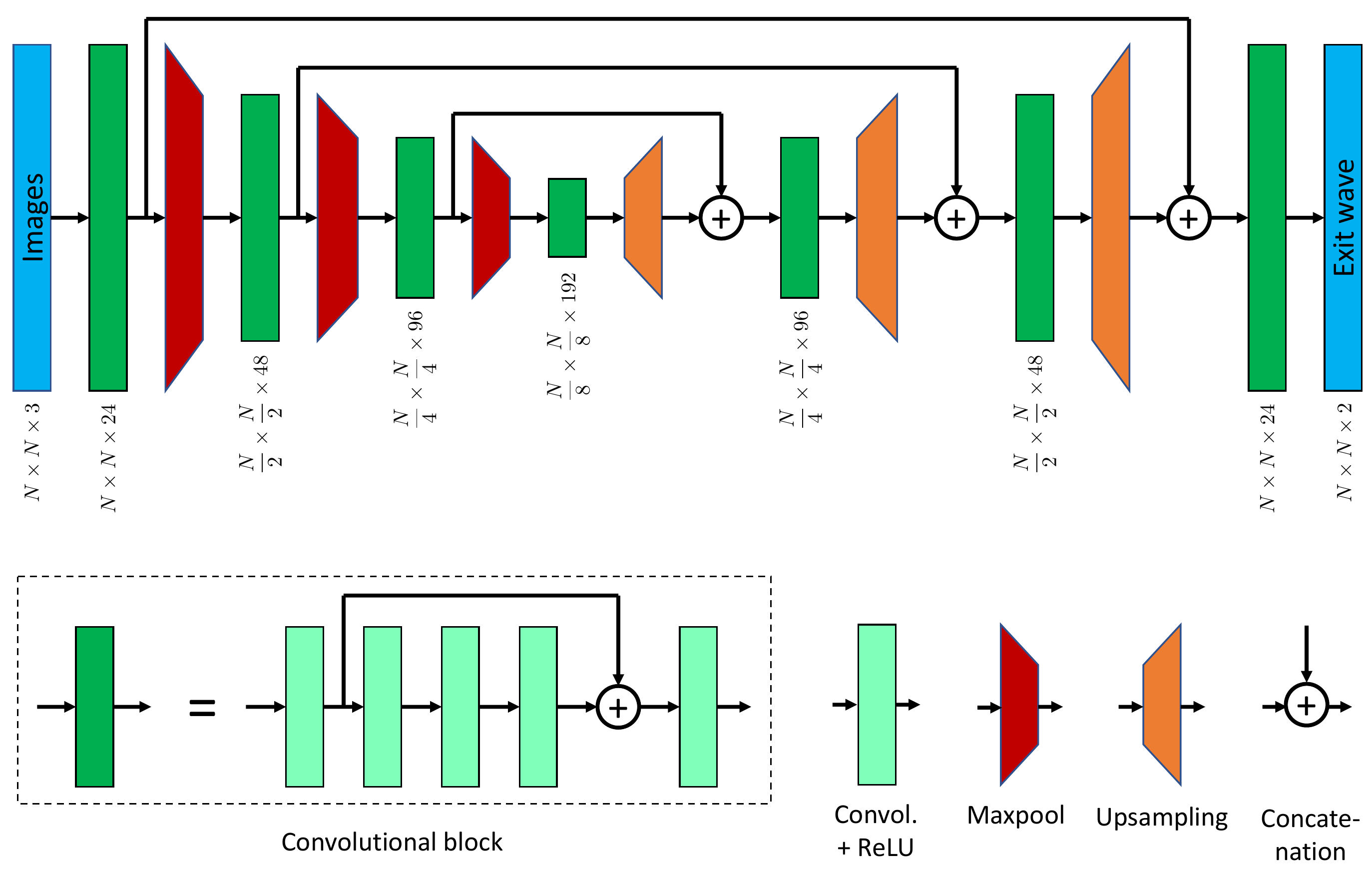}
  \caption{The architecture of the neural network.  Information flows
    from left to right.  The first part of the network, the ``encoding
    path'', consists of
    convolutional processing blocks alternated by downsampling layers
    using the MaxPool method.  The second part, the ``decoding path'',
    consists of convolutional blocks alternated by bilinear upsampling
    layers.  Long skip connections ensures that the original spatial
    information can be maintained.  Adapted from
    Madsen \emph{et al.} \cite{Madsen:2018ey}}
  \label{fig:arch}
\end{figure}

The neural network architecture is adapted from the one some of us
previously used \cite{Madsen:2018ey}.  It consists of a downsampling
(or ``encoding'') path, where convolutional blocks alternate with
downsampling layers, and an upsampling (or ``decoding'') path, where
the convolutional blocks alternate with upsampling layers (see
Fig.~\ref{fig:arch}).  The convolutional blocks consist of five
convolutional layers, with a short skip connection between the output
of the first layer and the input of the fifth.

The downsampling is done with conventional MaxPool operations.  Each
time the resolution is cut in half in a MaxPool operation, the number
of feature channels in the following convolutional block is doubled to
maintain the information flow in the network.  The upsampling is done
using bilinear interpolation, and the following convolution block has
the number of channels cut by a factor two.  After each upsampling,
information from the last layer with the same spatial resolution is
added from the downsampling path, this is done by concatenating the
channels, in contrast to Madsen \emph{et al.} where elementwise
addition was used.  The first layer in the convolutional blocks in
both paths will therefore have a different number of input channels
from what is stated in the figure.

Each convolutional layer uses a $3 \times 3$ convolutional kernel,
followed by a Parametric Leaky Rectifying Linear Unit.  During
hyperparameter optimization, we found that increasing the size of the
kernel to $5 \times 5$ or $7 \times 7$ did not improve the performance
of the network, nor did increasing the number of channels over the
numbers given in Fig.~\ref{fig:arch}. 

\subsection{Convergence of the multislice algorithm}
\label{sec:conv-multislice}

The multislice algorithm depends on a discretization of space, with
alternating interactions between the electron wave with the matter
within a slice of the material, and propagation of the wave from one
slice to another.  The finite slice thickness is an approximation. As the ``signal'' is
the change in the wave function from the unperturbed value of 1, we
define a measure of the relative change between wave functions
$\Psi_d$ and $\Psi_0$ as
\begin{equation}
  \label{eq:errormeasure}
  \Delta_d = {\int \left| \Psi_d(\vec r) - \Psi_0(\vec r)
    \right|^2 d\vec r \over \int \left| \Psi_o(\vec r) - 1 \right|^2 d\vec r}
\end{equation}

\begin{figure}
  \centering
  \includegraphics[width=0.95\linewidth]{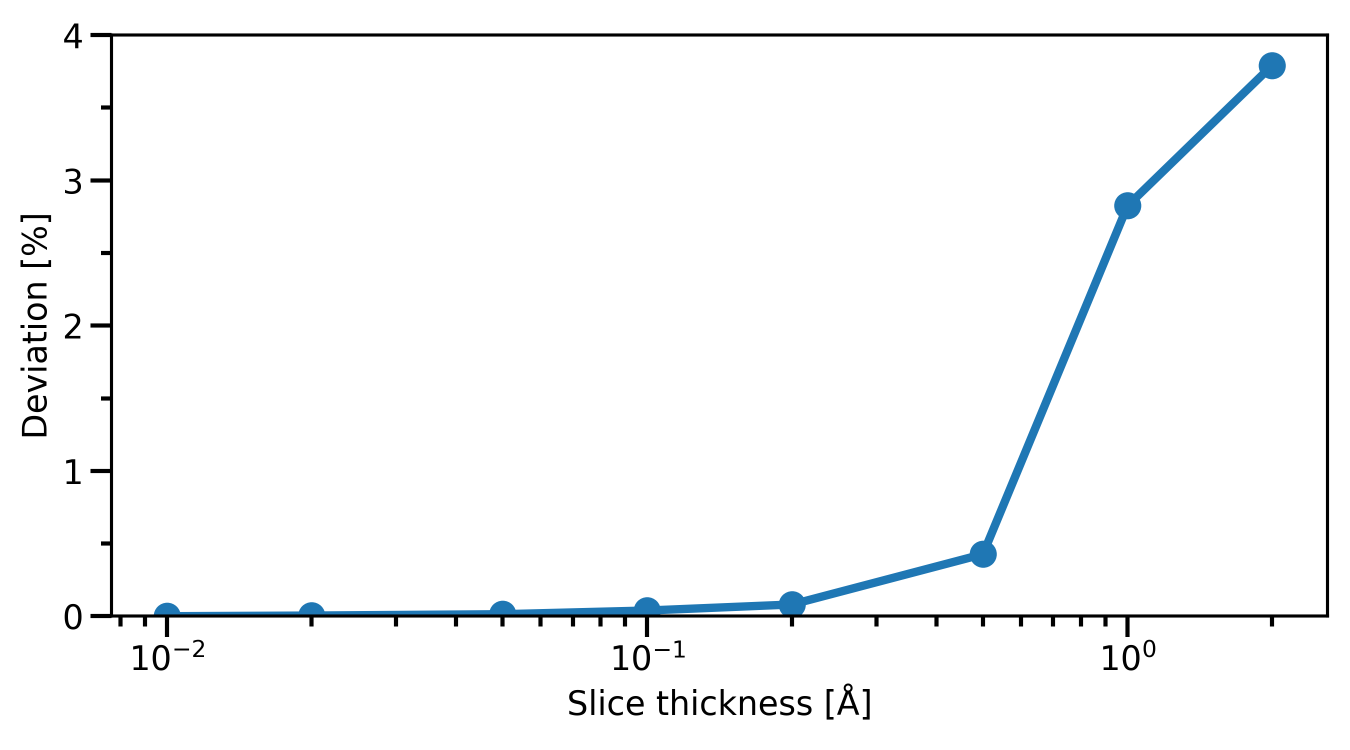}
  \caption{Convergence of the multislice algorithm with slice
    thickness $d$.  The deviation from the correct wave function is
    defined in equation \ref{eq:errormeasure}.}
  \label{fig:multisliceconvergence}
\end{figure}

In figure \ref{fig:multisliceconvergence}, we show the convergence of the calculated wave
function with slice thickness by plotting the difference between the
wave function calculated with various slice thicknesses $d$, using the one
calculated with $d = 0.01$ Å as the correct $\Psi_0$.

\subsection{Network training}
\label{sec:network-training}

\begin{figure}
  \centering
    \includegraphics[width=\linewidth]{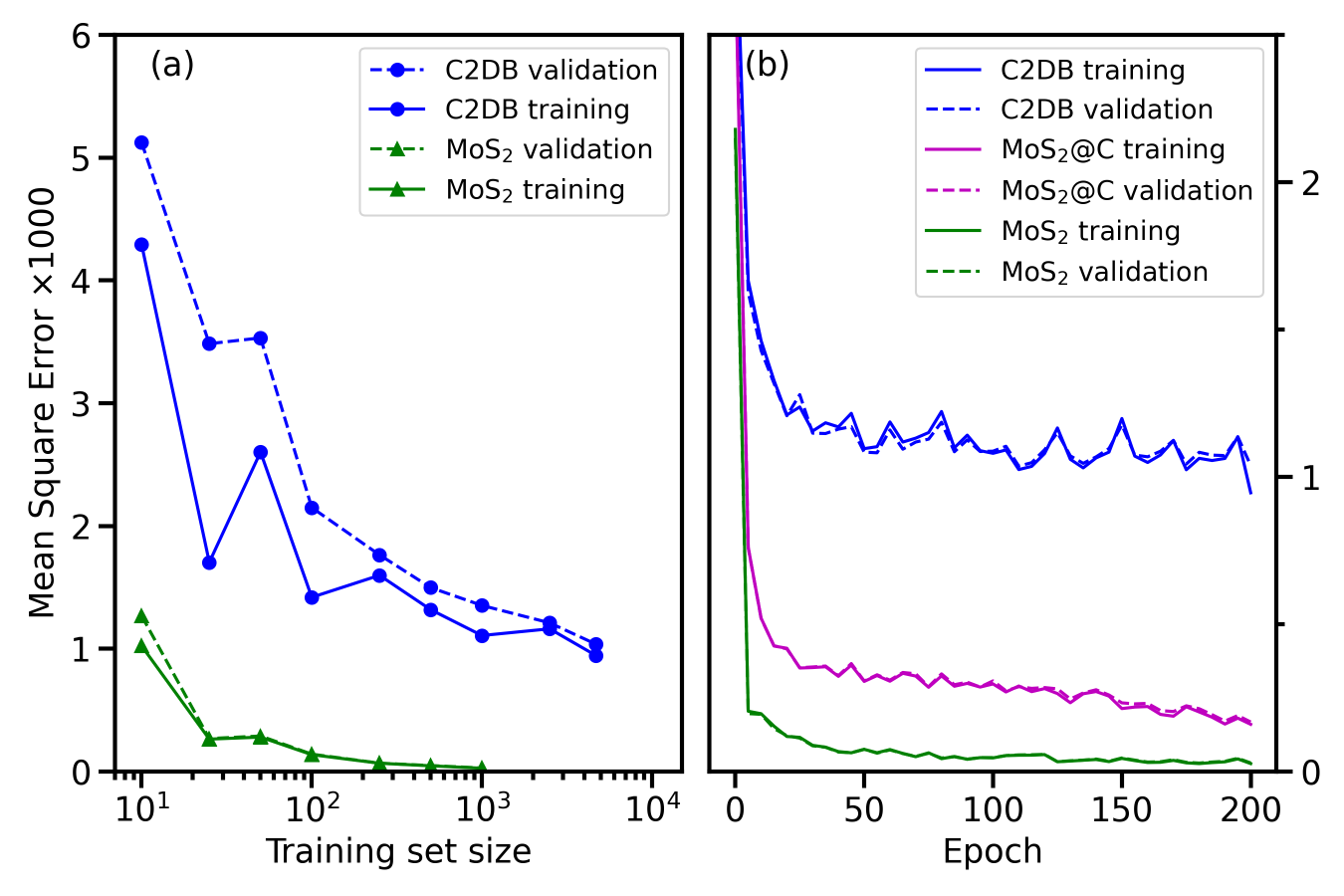}
    \caption{Learning curves: (a) Training and validation loss as a
      function of training set size for the C2DB and the unsupported
      MoS$_2$ data sets.  The validation sets were kept constant, but
      only a fraction of the training sets were used to train the
      networks.  (b) Training and validation loss as a function of
      epoch number (calculated for every 5 epochs).  Note that the
      validation and training curves are almost identical.}
  \label{fig:trainingcurves}
\end{figure}
As mentioned in the main text, training is done using the Mean Square
Error (MSE) loss function, and the RMSprop algorithm.  To check for
overfitting, we artificially limited the size of the training set, and
calculated both training and validation losses as a function of
training set size, this is shown in Fig.~\ref{fig:trainingcurves}(a).  As
no increase in the validation loss is seen for the largest training
sets, we conclude that overfitting is not an issue.  Figure
~\ref{fig:trainingcurves}(b) show the same losses as a function of
epoch number, again no overfitting is seen, but perhaps the network
trained on MoS2 supported by graphene could have been even better with
a bit more training.

\subsection{Gaussian smearing of the wave function.}
\label{sec:gauss-smear-wave}

\begin{figure}
  \centering
  \includegraphics[width=\linewidth]{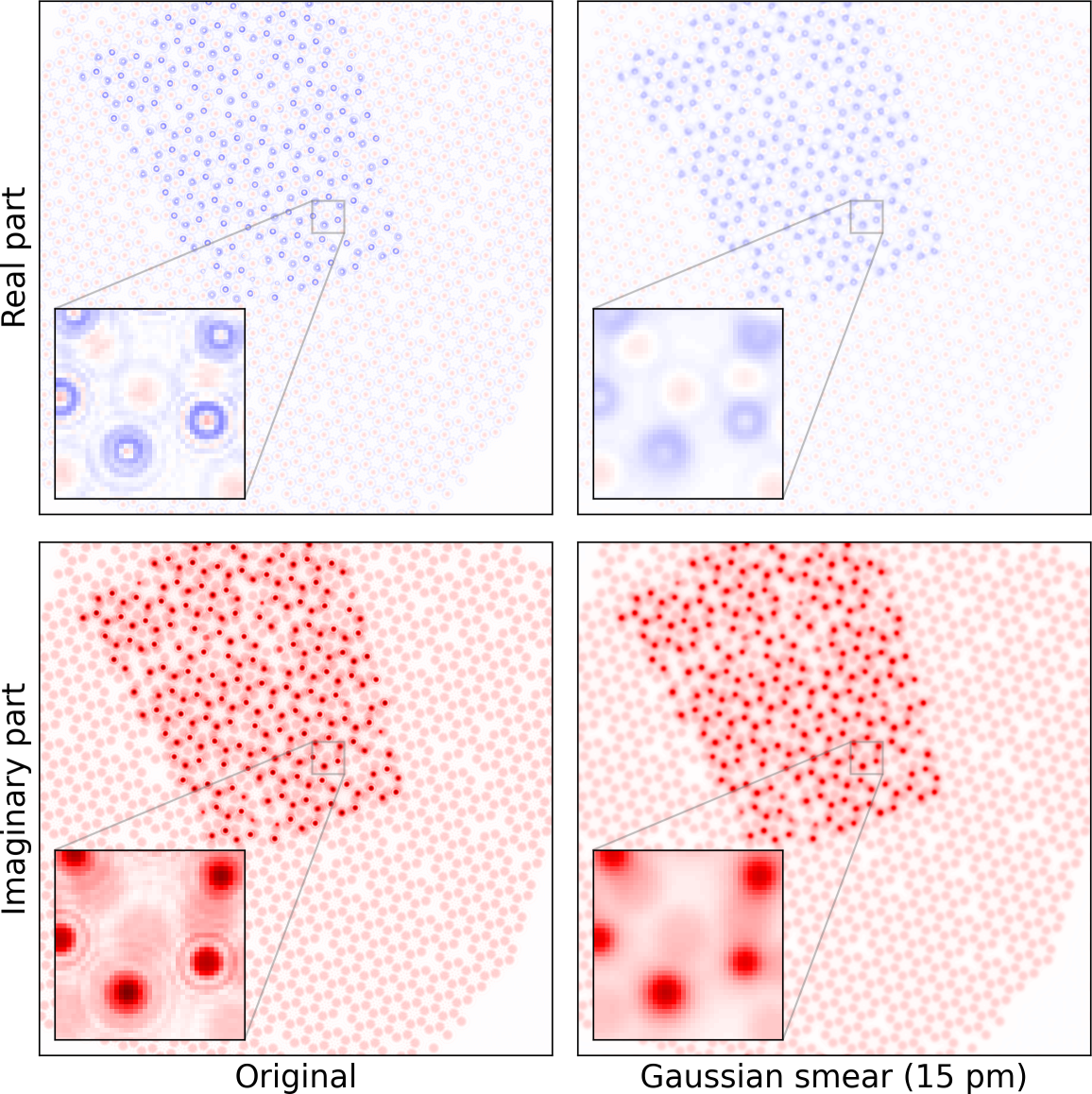}
  \caption{Effect of Gaussian smearing on the wave function of the
    MoS$_2$ nanoparticle also shown in Fig.~\protect\ref{fig:MoS2sup} of the
    main text.  Left column shows the real and imaginary part of the
    exit wave without smearing, the right column shows them after
    Gaussian smearing with a width of 20 pm.  The ring-shaped
    structures are suppressed, allowing the network to describe the
    main peak better.}
  \label{fig:smearing}
\end{figure}

The effect of Gaussian smearing of the wave function is shown in
Fig.~\ref{fig:smearing}.

\subsection{Examples of good and bad predictions.}
\label{sec:examples-good-bad}

\begin{figure}
  \centering
  \includegraphics[width=\linewidth]{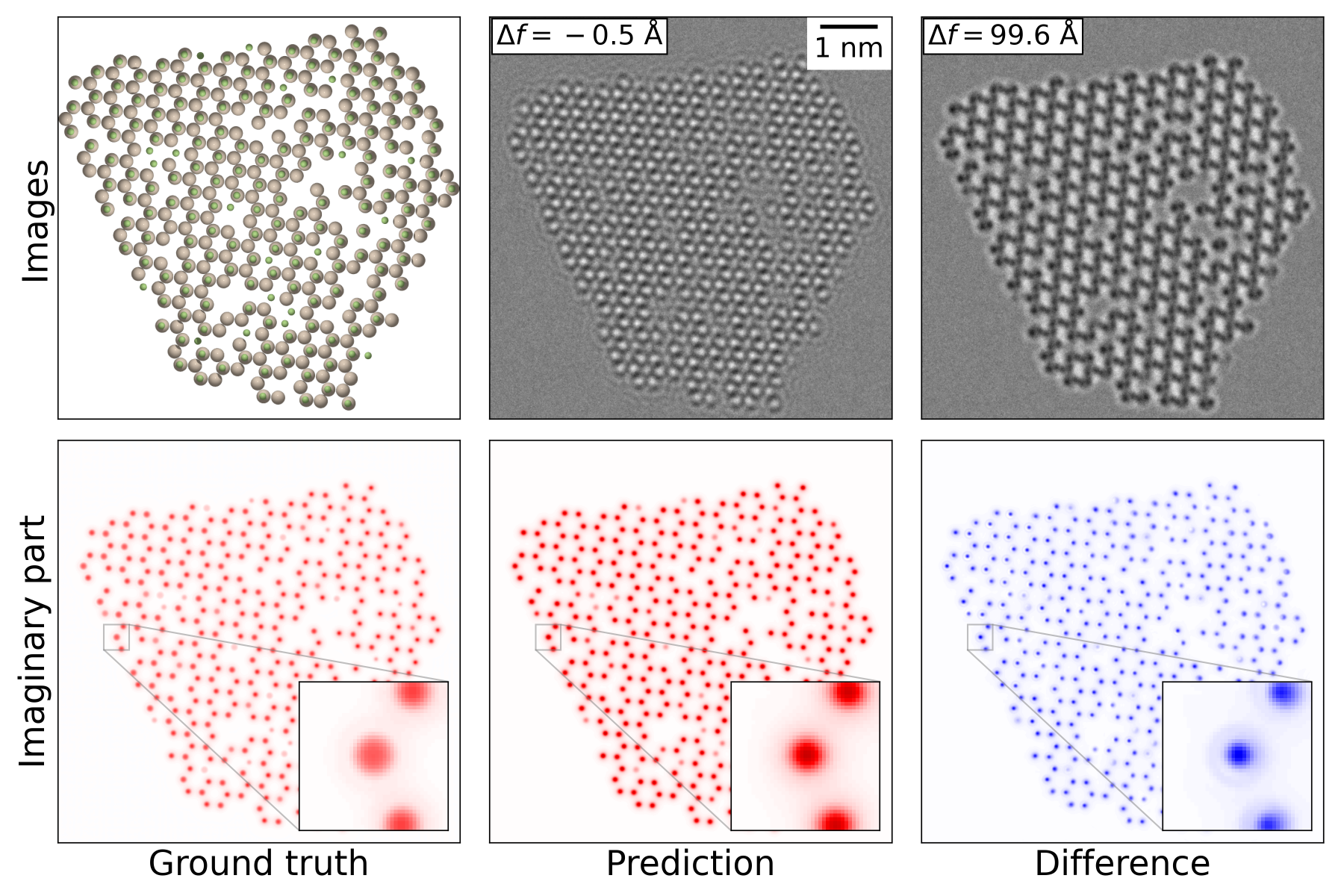}
  \caption{The 5\% image from the C2DB dataset.  Five percent of the
    images are this bad or worse.  The material is SiF (Silicon
    fluoride), and the main error is a systematic overestimation
    of the intensity of the peaks  The worst placed peak is displaced
    by 11 pm (1 pixel).}
  \label{fig:c2db-5p}
\end{figure}

\begin{figure}
  \centering
  \includegraphics[width=\linewidth]{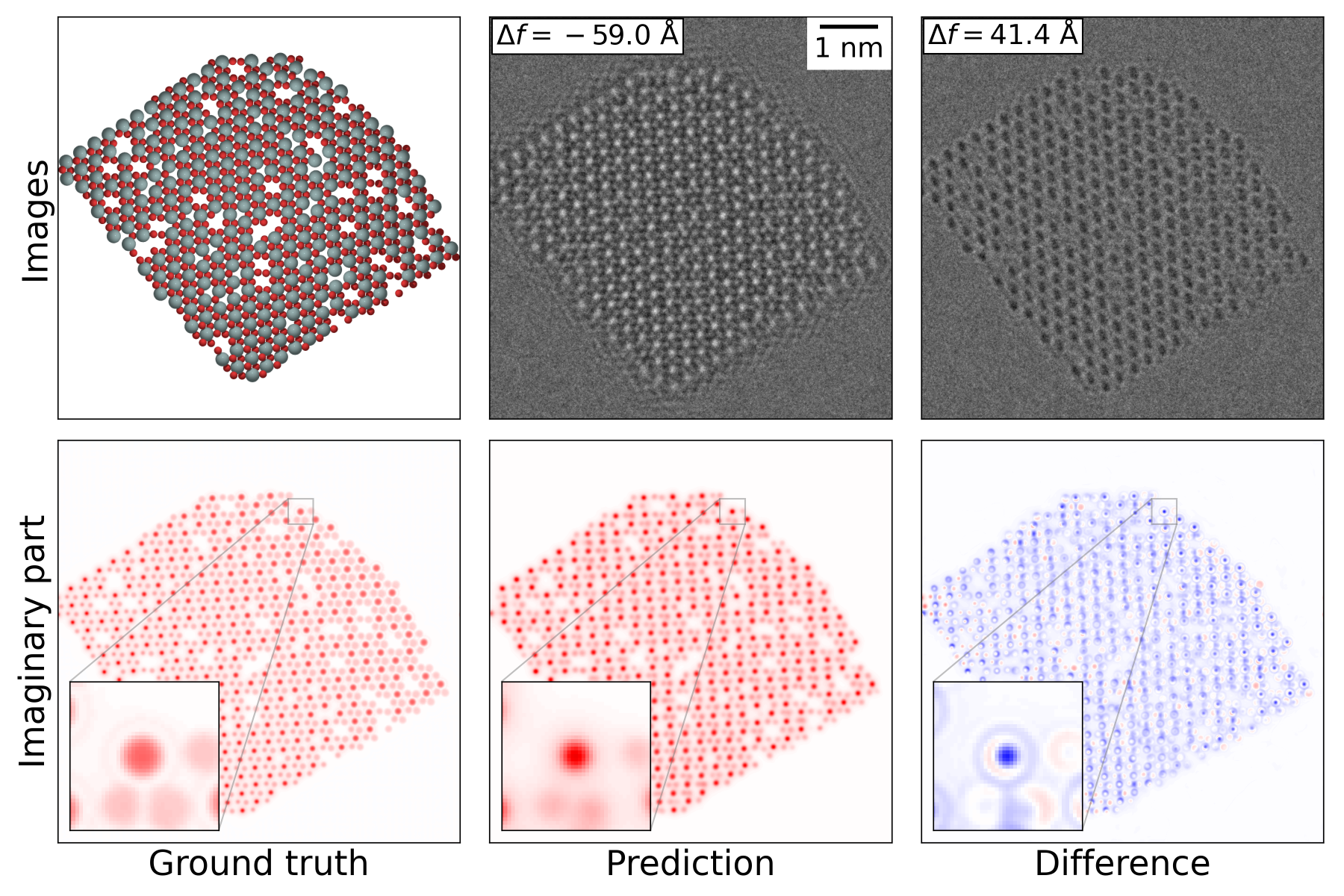}
  \caption{The 25\% image from the C2DB dataset.  The material is
    GeO$_2$ (Germanium oxide), and the main error is an
    overestimation of the peaks relating to the Ge atoms. The worst placed peak is displaced
    by 14 pm (1.3 pixels).}
  \label{fig:c2db-25p}
\end{figure}

\begin{figure}
  \centering
  \includegraphics[width=\linewidth]{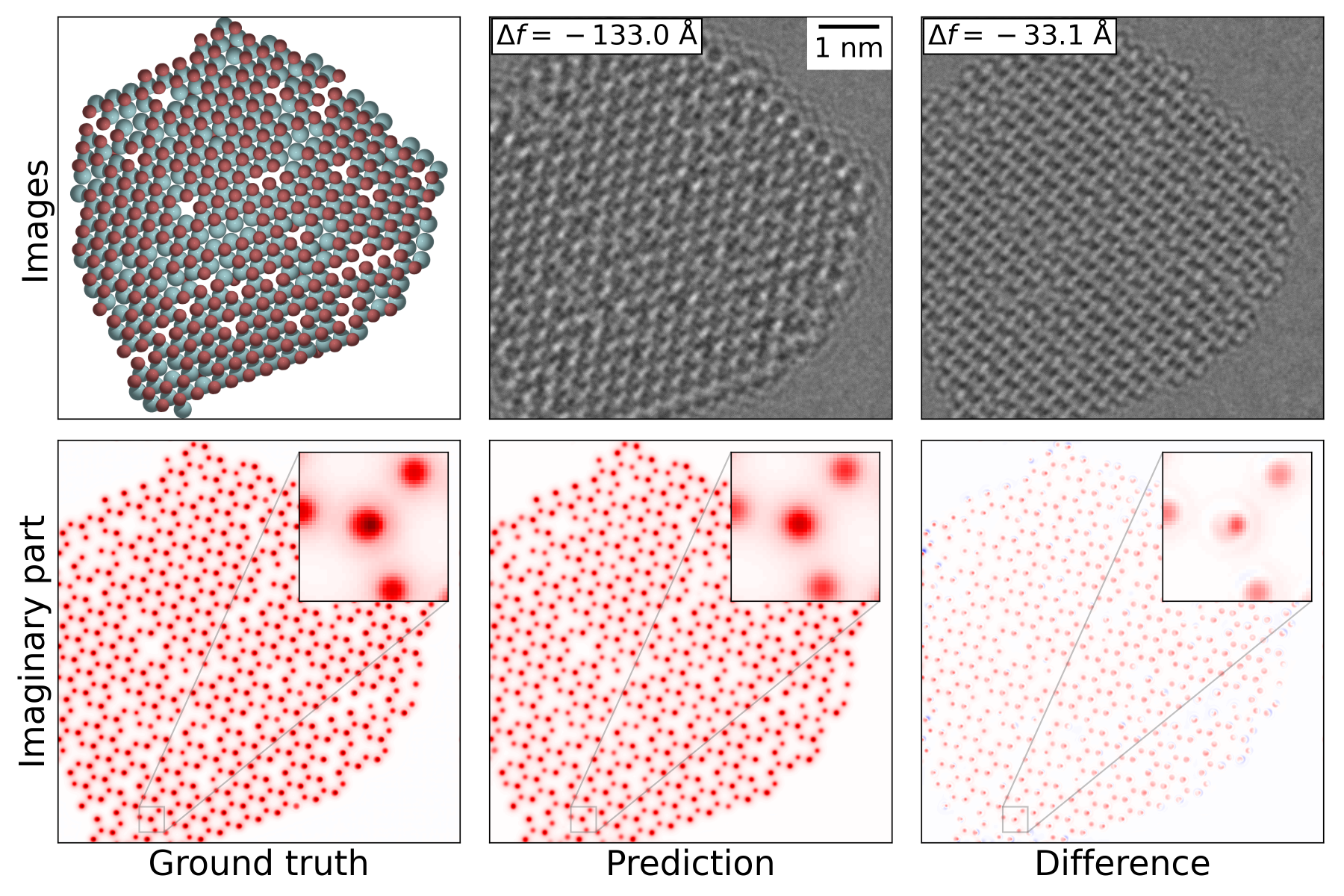}
  \caption{The 75\% image from the C2DB dataset.  The network slightly
    underestimates the intensity of the peaks in this NbBr$_2$
    (Niobium bromide). The worst placed peak is displaced
    by 8 pm (0.8 pixels).}
  \label{fig:c2db-75p}
\end{figure}

\begin{figure}
  \centering
  \includegraphics[width=\linewidth]{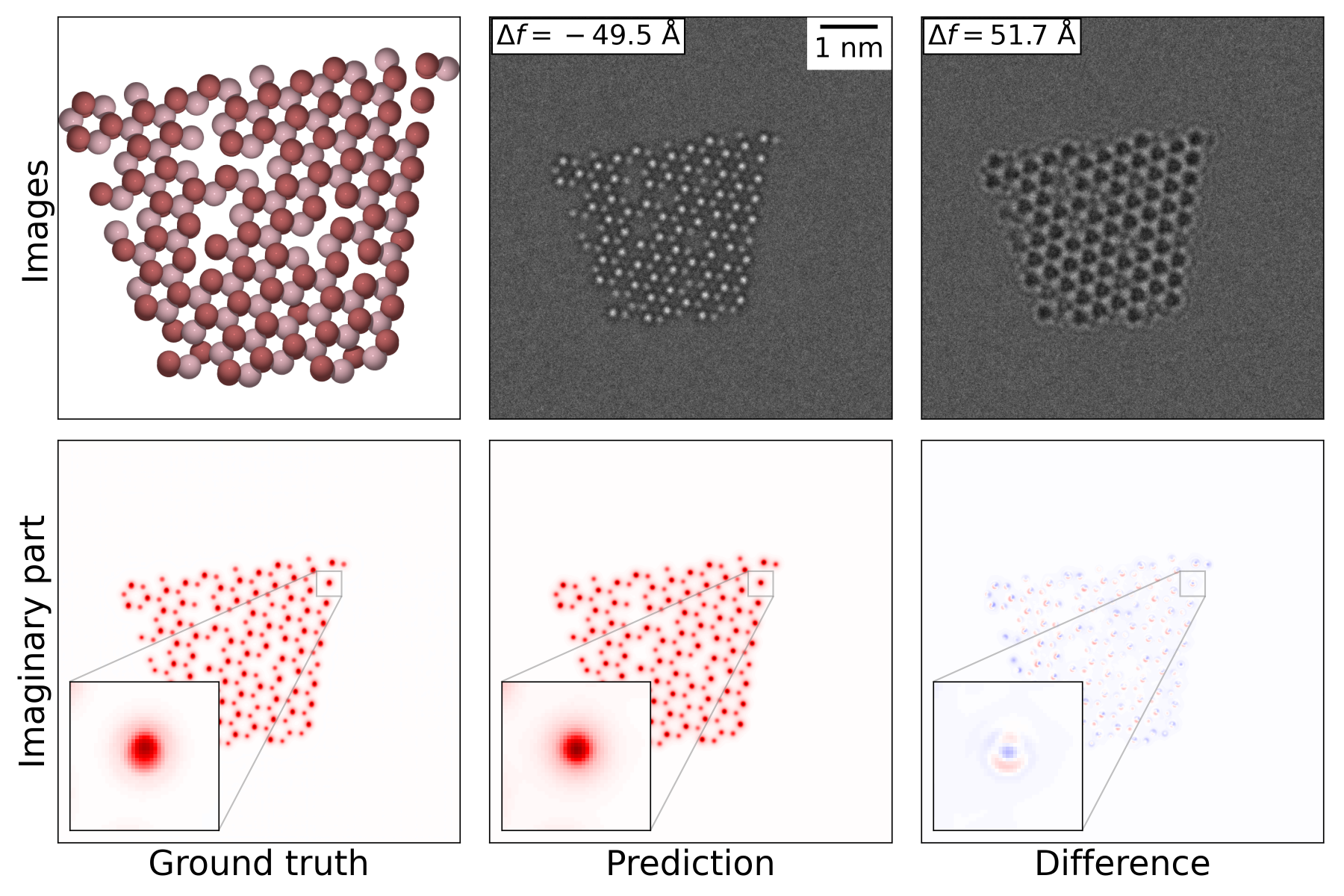}
  \caption{The 95\% image from the C2DB dataset, only five percent of
    preditions are better than this.  The system is CoBr$_2$ (Cobalt
    bromide), and the exit wave is reconstructed almost perfectly,
    with a slight misplacement of some peaks by up to 9 pm (0.8 pixels).}
  \label{fig:c2db-95p}
\end{figure}

In Figures \ref{fig:c2db-5p} to \ref{fig:c2db-95p}, we show examples
of the predictions of the network on the C2DB dataset, selected by ranking the results
obtained on the validation set, and showing images with 5\% percentile
rank, 25\%, 75\% and 95\%, meaning that 5\% (etc) of the images
are worse than the image shown.  The 50\% images are shown in the main
text.

\begin{figure}
  \centering
  \includegraphics[width=\linewidth]{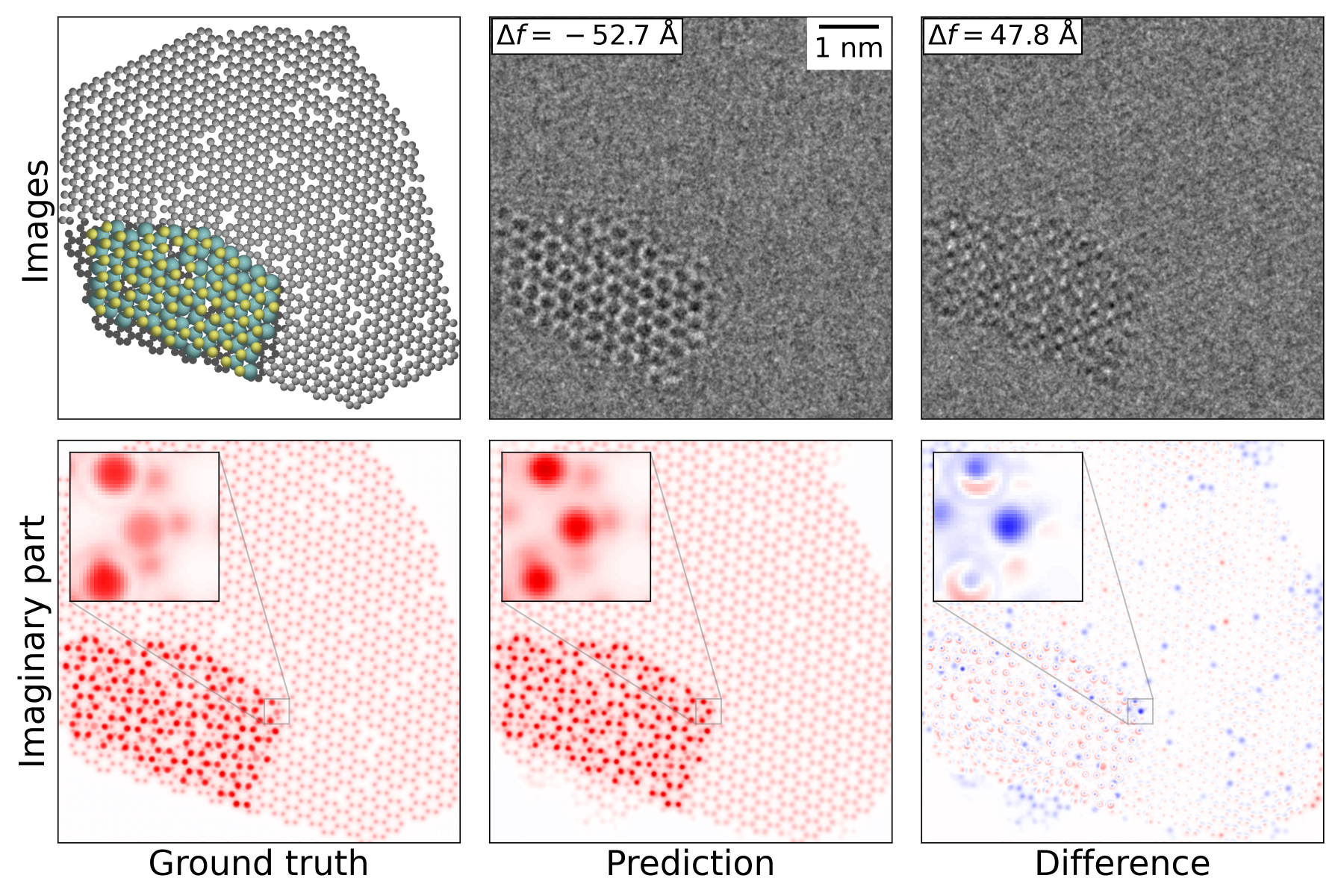}
  \caption{The 5\% image from the MoS$_2$ dataset.  In spite of the
    bad contrast in the images, the network mostly resolves the structure.
    It
    misses a few sulphur vacancies in the MoS$_2$ layer and a
    significant number of vacancies in the graphene substrate.  The
    largest error in positioning Mo or S atoms is 16 pm (1.5 pixels).}
  \label{fig:mos2-5p}
\end{figure}

\begin{figure}
  \centering
  \includegraphics[width=\linewidth]{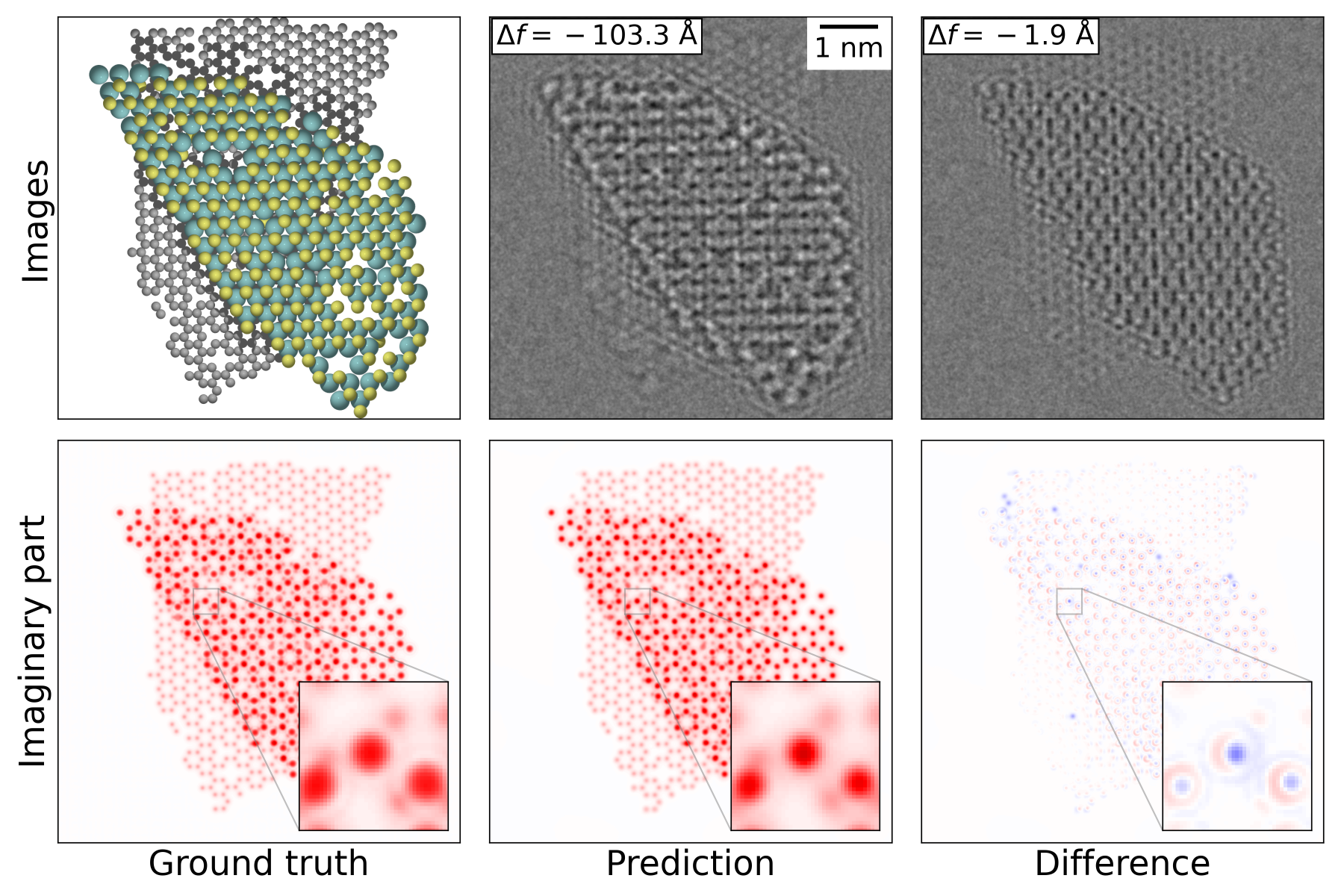}
  \caption{The 25\% image from the MoS$_2$ dataset.  All defects and atoms are
    correctly localized, the only error the network produces is in the
    shape of the peaks. The largest error in positioning is 9.5 pm
    (below 1 pixel).}
  \label{fig:mos2-25p}
\end{figure}

\begin{figure}
  \centering
  \includegraphics[width=\linewidth]{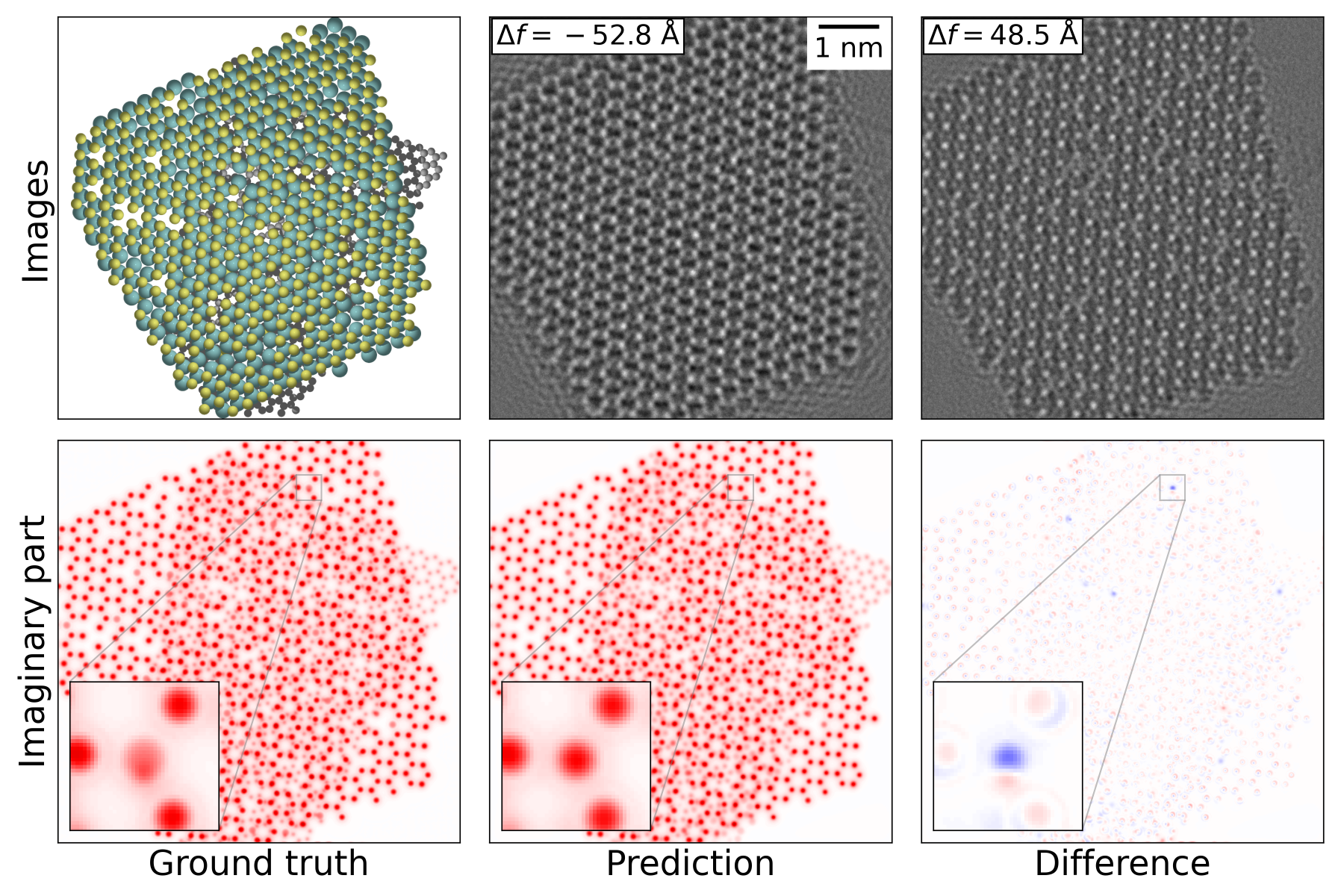}
  \caption{The 75\% image from the MoS$_2$ dataset.  Here the network
    mistakes a single sulphur vacancy that coincides with a carbon
    support atom for a sulphur dimer above a carbon vacancy.}
  \label{fig:mos2-75p}
\end{figure}

\begin{figure}
  \centering
  \includegraphics[width=\linewidth]{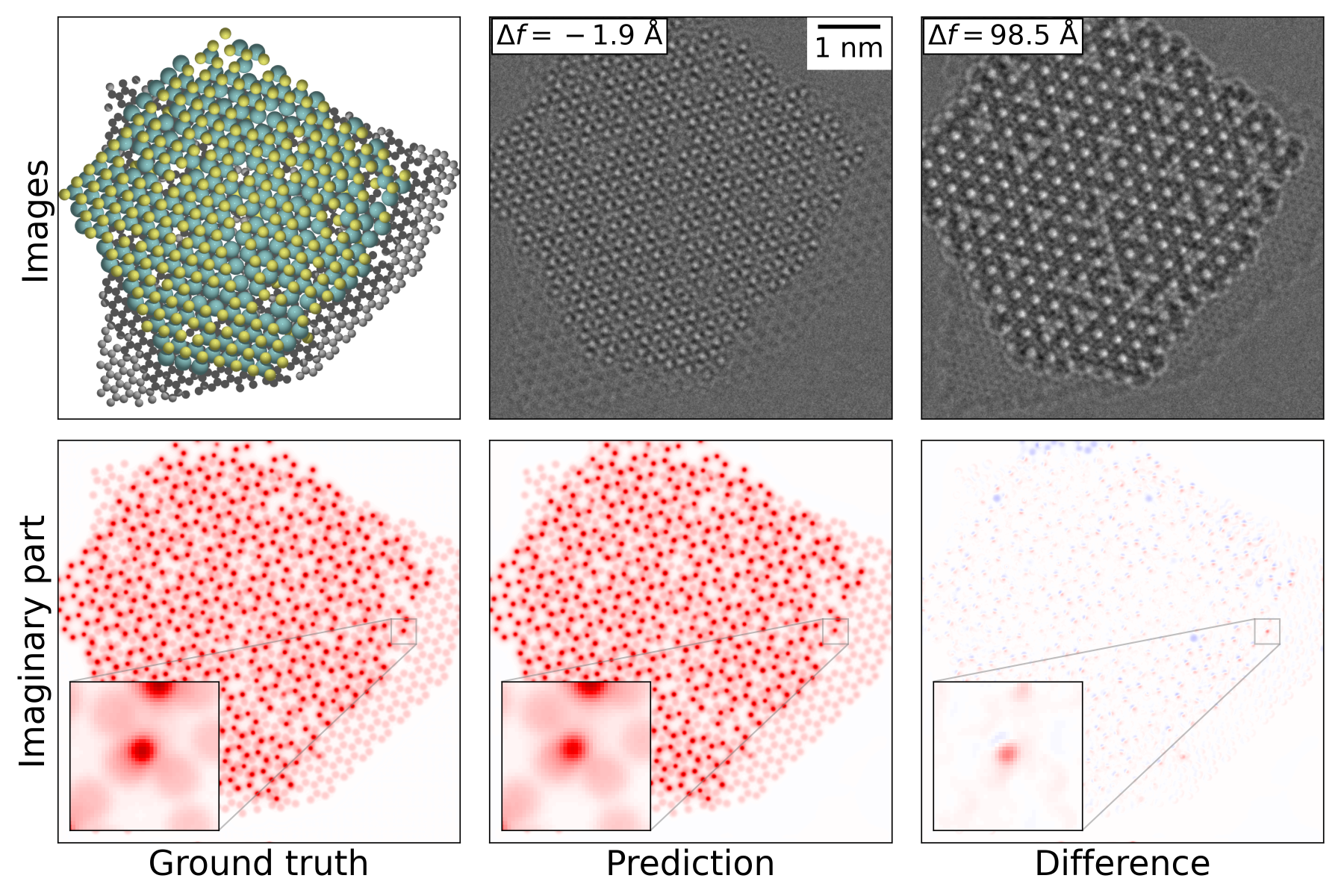}
  \caption{The 95\% image from the MoS$_2$ dataset.  The exit wave of
    this unsupported particle is almost perfectly reproduced.}
  \label{fig:mos2-95p}
\end{figure}

In the same way, we show the same percentiles for the data set of
supported MoS$_2$ (Figs.~\ref{fig:mos2-5p} to \ref{fig:mos2-95p}).  

The performance measure used to select the images is based on the mean
square error (MSE) of the reconstructed wave function.  However, the
areas of the nanoparticles vary substantially, and the network finds
it very easy to reproduce the value 1 in the vacuum, leading to an MSE
that is just as much determined by the area of the nanoparticle as by
the quality of the prediction.  We therefore divide by the total
signal in the exit wave, as that is proportional to the area, and
select the images by the quantity defined in
Eq.~(\ref{eq:errormeasure}) where $\Psi_{\text{d}}$ is now the
predicted wave function and $\Psi_{\text{0}}$ is the ground
truth.

\subsection{Comparison with Gerchberg-Saxton exit wave
  reconstruction.}
\label{sec:comp-with-gerchb}

\begin{figure}
  \centering
   \includegraphics[width=\linewidth]{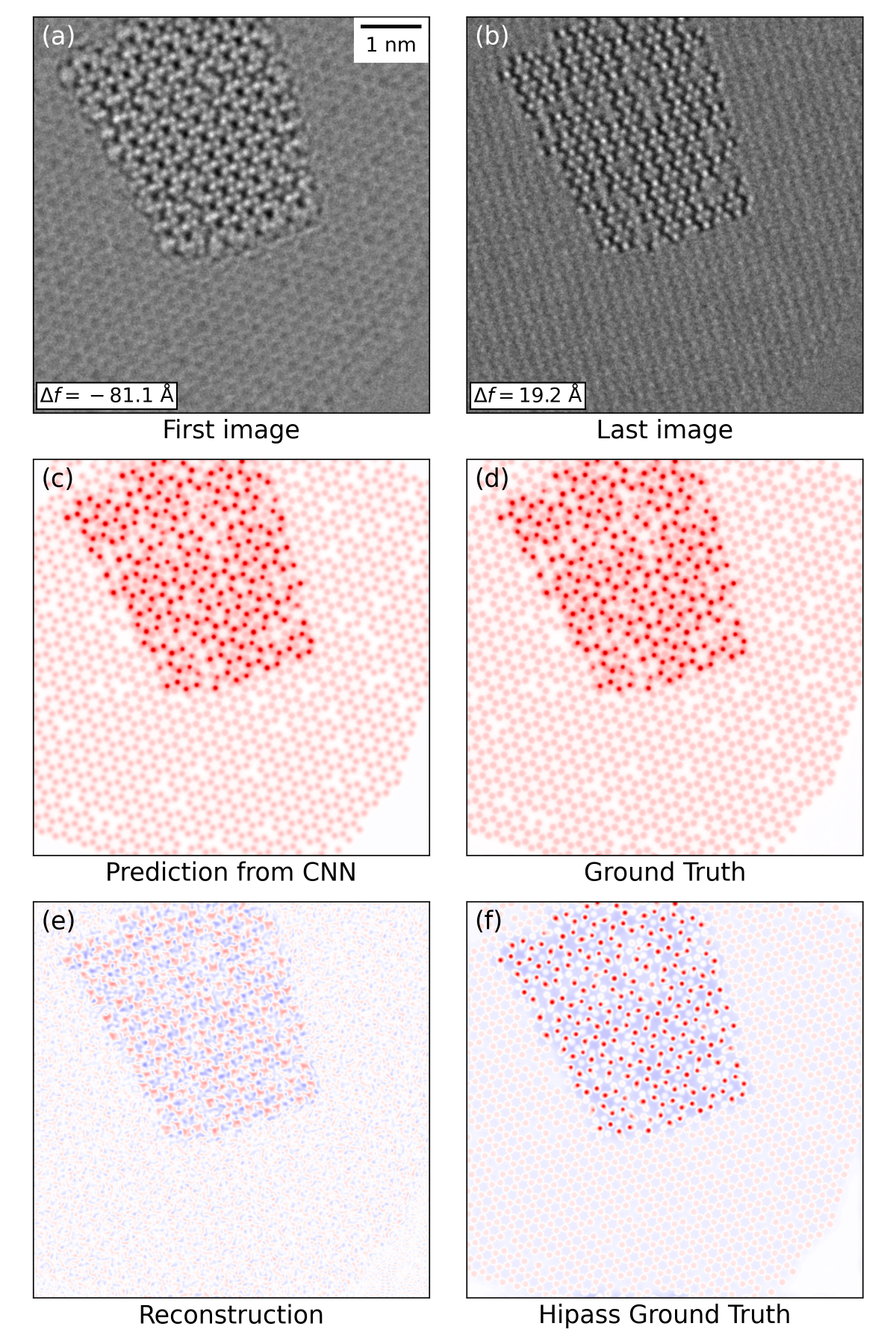}
   \caption{Comparing the neural network with a traditional algorithm
     for exit wave reconstruction for the 50 percentile supported
     MoS$_2$ system (shown in the main text, Fig.~\ref{fig:MoS2sup}). (a) and (b) The first and last
     images in the image series.  (c) The exit wave reconstructed by
     the network.  (d) The ground truth (correct) exit wave.  (e) The
     exit wave reconstructed by the Gerchberg-Saxton algorithm.  The
     large deviations are due to the long wavelength part of the exit
     wave not being reconstructed.  (f) The ground truth wave function
     with the longest wavelengths removed.}
  \label{fig:reconst50p}
\end{figure}

\begin{figure}
  \centering
   \includegraphics[width=\linewidth]{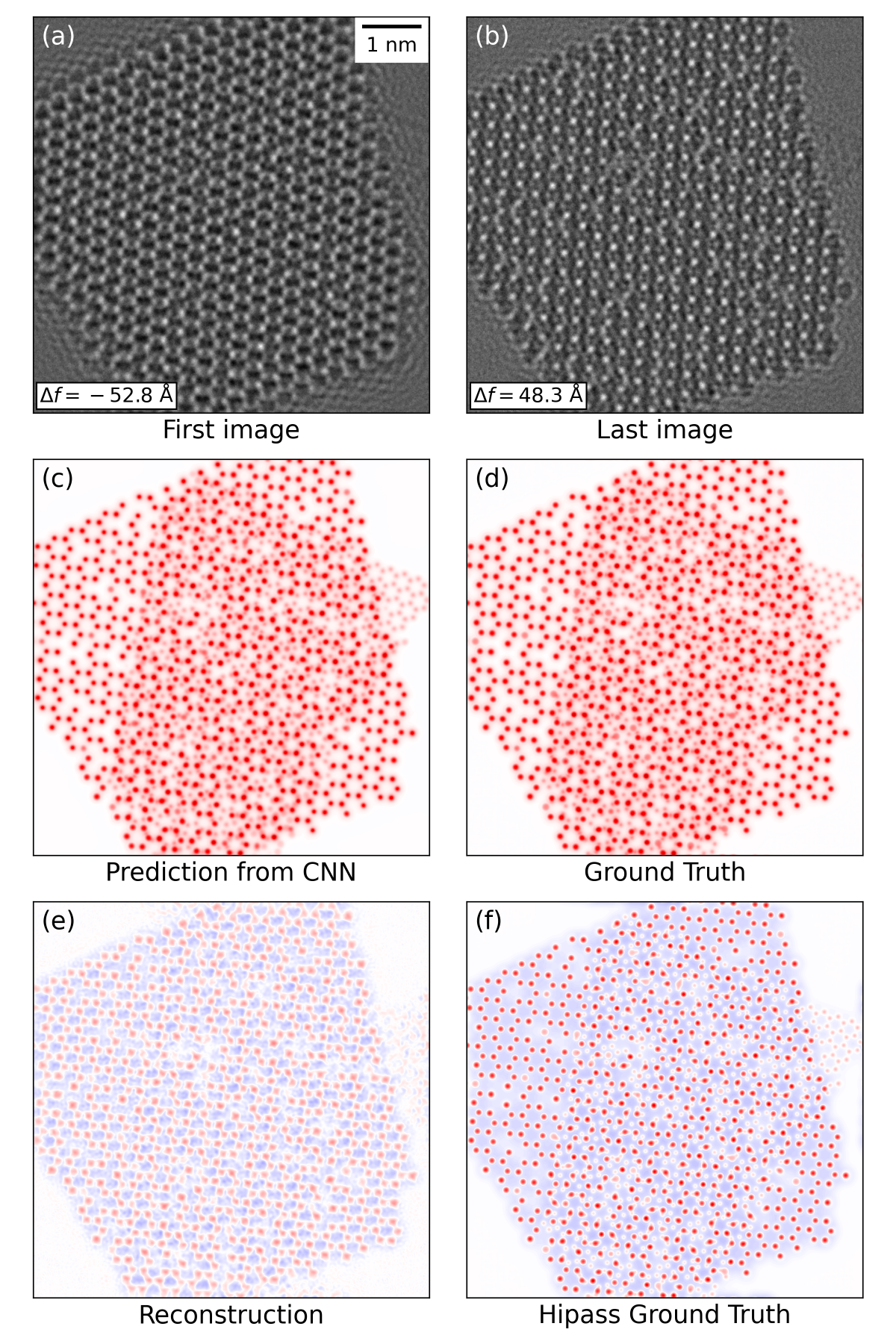}
   \caption{Comparing the neural network with a Gerchberg-Saxton algorithm
     for exit wave reconstruction for the 75 percentile supported
     MoS$_2$ system (shown in Fig.~\ref{fig:mos2-75p}.}
  \label{fig:reconst75p}
\end{figure}

Three systems were reconstructed using the Gerschberg-Saxton
algorithm, as described in the main text.  The reconstructed wave for
the 25-percentile system i Figure \ref{fig:mos2-25p} is shown in the
main text.  The reconstructed waves for the 50-percentile and
75-percentile systems (Figures \ref{fig:MoS2sup} (main text) and
\ref{fig:mos2-75p}) are shown in Figures \ref{fig:reconst50p} and
\ref{fig:reconst75p}.  We see that the 50-percentile system has been
harder to reconstruct with the Gerschberg-Saxton algorithm than the
other two, presumably because both 2-fold astigmatism and coma of the
simulated are
near the upper limits in their respective distributions (main text, Table \ref{tab:parameters}).

\subsection{Computational ressources}
\label{sec:comp-ress}

The computational ressources necessary for a project like this are
relatively modest.  We here list the amount of computational
ressources used for the three phases of the project: Multislice
simulations of the exit wave, image generation, and neural network
training.

The computations were done on the Niflheim supercomputer cluster at
DTU, but always on single servers within the cluster.
\begin{itemize}
\item The multislice simulations were done on servers with two 12-core
  Intel Broadwell processors (24 cores in total) running at 2.20 GHz,
  installed in 2017.
\item The image generations were done on servers with two 20-core
  Intel Skylake processors (40 cores in total) running at 2.40 GHz,
  installed in 2019.  The image generation could easily have been done
  on the above-mentioned servers instead.
\item The network training was done on a single Nvidia RTX 3090 GPU on
  a shared multi-gpu server.
\end{itemize}
The amount of time used for the three phases for the three different
kinds of systems are shown in Table \ref{tab:ressources}.

\begin{table}
  \centering
  \begin{tabular}{l|lll}
    \hline\hline
    System & Multislice & Image generation & Training \\
    \hline
    MoS$_2$ & 2 h 45 min & 1 h 44 min & 25 h \\
    MoS$_2$@C & 3 h 3 min & 1 h 48 min & 24 h \\
    C2DB & 14 h 2 min & 8 h 35 min & 88 h\\
    \hline\hline
  \end{tabular}
  \caption{Computational ressources}
  \label{tab:ressources}
\end{table}



\end{document}